\documentclass[aps,nofootinbib,superscriptaddress,showpacs,preprintnumbers,10pt,prd,nofootinbibt,onecolumn]{revtex4-2}
\usepackage{booktabs}
\pdfoutput=1

\usepackage[utf8]{inputenc}
\usepackage[T1]{fontenc}
\usepackage{amsmath}
\usepackage{amssymb}
\usepackage{graphicx}
\usepackage{caption}
\usepackage{subcaption}
\usepackage[most]{tcolorbox}
\usepackage{mathtools}
\usepackage{epsfig}
\usepackage{multirow}
\usepackage{eurosym}
\usepackage{dcolumn}
\usepackage{bm}
\usepackage{enumerate}
\usepackage{float}
\usepackage{epstopdf}
\usepackage{amsmath}
\usepackage{bm}
\usepackage{amsfonts}
\usepackage{amssymb}
\usepackage{graphicx}
\usepackage{alphalph,mathtools}
\usepackage{etoolbox}
\usepackage{color}
\usepackage{booktabs}
\usepackage{footnote}
\usepackage{makecell,tabularx}
\usepackage[leftcaption]{sidecap}
\usepackage{amssymb}
\usepackage{graphicx}
\usepackage{booktabs,dcolumn,caption}
\usepackage{rotating}
\usepackage{amsthm}
\usepackage{tikz}
\usepackage{caption}
\usepackage{subcaption}
\usepackage[most]{tcolorbox}

\usepackage{hyperref}
\hypersetup{colorlinks,citecolor=blue}
\hypersetup{colorlinks=true,linkcolor=magenta,filecolor=green, urlcolor=blue}

\def\be{\begin{equation}}
	\def\ee{\end{equation}}
\def\bea{\begin{eqnarray}}
	\def\eea{\end{eqnarray}}

\usepackage{xcolor}
\def\beq{\begin{eqnarray}}  
	\def\eeq{\end{eqnarray}}

\begin{document}
	
	\title{  Traversable Wormhole Solutions in massive  $F(T)$ 
gravity }
	
	\author{A. Landry} %\orcidlink{0000-0002-1046-439X}}
\email{a.landry@dal.ca}
\affiliation{Department of Mathematics and Statistics, Dalhousie University, Halifax, Nova Scotia, Canada, B3H 3J5}

\author{Y. Sekhmani} %\orcidlink{0000-0001-7448-4579}}
\email{sekhmaniyassine@gmail.com}
\affiliation{Center for Theoretical Physics, Khazar University, 41 Mehseti Street, Baku, AZ1096, Azerbaijan.}
\affiliation{Centre for Research Impact \& Outcome, Chitkara University Institute of Engineering and Technology, Chitkara University, Rajpura, 140401, Punjab, India.}

\author{S. K. Maurya}% \orcidlink{0000-0003-0261-7234}}
\email{sunil@unizwa.edu.om}
\affiliation{Department of Mathematical and Physical Sciences, College of Arts and Sciences, University of Nizwa, Nizwa 616, Sultanate of Oman}

\author{A. Ali}
\email{akali@kku.edu.sa}
\affiliation{Department of Mathematics, College of Sciences, King Khalid University, Abha 61413, Saudi Arabia}

\author{E.N. Saridakis}
\email{msaridak@noa.gr}
\affiliation{The National Observatory of Athens, Lofos Nymfon 11852, Greece}
\affiliation{Departamento de Matem\'{a}ticas, Universidad Cat\'{o}lica del Norte, Avda. Angamos 0610, Casilla 1280, Antofagasta, Chile}
\affiliation{CAS Key Laboratory for Research in Galaxies and Cosmology, School 
of Astronomy and Space Science,
University of Science and Technology of China, Hefei 230026, China}

%\section*{Abstract}
\begin{abstract}
We study traversable wormhole geometries in an $F(T)$ teleparallel framework augmented by a perturbative de Rham–Gabadadze–Tolley (dRGT) graviton-mass term. Adopting the static, spherically symmetric Morris–Thorne ansatz, we derive the field and conservation equations and decompose the effective energy–momentum tensor into torsional and massive contributions. Focusing on three representative redshift profiles, namely, constant, logarithmic and power-law, together with two realisations of the massive sector (the general case and a uniform-pressure specialisation), we construct exact, horizonless solutions that satisfy the Morris–Thorne flaring-out condition and are asymptotically flat. The effective matter sector either respects the standard energy conditions or only mildly violates them within controlled parameter ranges. Crucially, the dRGT term supplies an additional anisotropic pressure that can sustain the throat without invoking explicitly exotic matter; in the vanishing-mass limit the configurations reduce smoothly to standard $F(T)$ wormholes, confirming the internal consistency of the framework.

\end{abstract}

\maketitle

\section{Introduction}\label{sec1}

The study of modified theories of gravity has been motivated by the need to 
understand both the early- and late-time dynamics of the Universe, as well as to 
address theoretical issues arising in the ultraviolet regime of General 
Relativity (GR). Among the large number of alternative formulations 
\cite{CANTATA:2021asi}, the teleparallel approach provides a geometrical 
framework in which gravitation is attributed to torsion rather than curvature 
\cite{Aldrovandi_Pereira2013,Cai:2015emx,Bahamonde:2021gfp,Krssak:2018ywd,
Coley:2019zld,Lucas_Obukhov_Pereira2009,
Krssak_Pereira2015,hohmannfq}. In this context, the fundamental dynamical 
variable is the tetrad field, and the corresponding Weitzenböck connection 
ensures vanishing curvature while retaining non-zero torsion 
\cite{Aldrovandi_Pereira2013}. The simplest 
teleparallel theory, namely the Teleparallel Equivalent of General Relativity 
(TEGR), reproduces Einstein’s field equations (FEs) but through the torsion scalar $T$ 
instead of the Ricci scalar $R$ \cite{chinea1988symmetries,estabrook1996moving,
papadopoulos2012locally,olver1995equivalence,MCH}. Extending TEGR to a general 
function $F(T)$ leads to the $F(T)$ class of modified gravities, which can 
account for inflationary and dark-energy behaviors without introducing an 
explicit cosmological constant 
\cite{Chen:2010va,Ferraro:2008ey,Krssak:2015oua,golov3}. In 
these classes of theories, any 
tetrad/spin-connection pair will be solution of Lie derivatives and isotropy 
conditions \cite{chinea1988symmetries,estabrook1996moving,
papadopoulos2012locally,olver1995equivalence,MCH}. Finally, one can extend the 
above considerations 
in other geometric modifications of gravity, such as  $F(T,B)$-type, 
$F(Q)$-type, scalar-torsion theories, and so on~\cite{kayashi,beltranngr,bahamondengr,heisenberg1,heisenberg2,faithman1,
jimeneztrinity,nakayama,ftqgravity,frqspecial,frtspecial,frttheory,
Saridakis:2019qwt,Anagnostopoulos:2021ydo,Basilakos:2025olm}.

Another class of modified theories of gravity is based on the 
possibility that the graviton possesses a small but finite mass. Massive gravity theories introduce additional degrees of freedom in the metric sector and can effectively reproduce self-accelerating solutions without a cosmological constant. The most consistent realization is the de Rham-Gabadadze-Tolley (dRGT) theory, which eliminates the Boulware-Deser ghost through a specific structure of the potential term \cite{massivefrgravity1,derham1,derham2,derham3,derham4,derham5,
derham6}.  The weak or linearized version of the dRGT potential can be consistently 
incorporated into other modified gravity frameworks, allowing one to investigate combined effects of torsion and graviton mass at astrophysical scales.

A considerable amount of work has been devoted to black-hole and wormhole 
solutions in both curvature- and torsion-based modified theories of gravity 
\cite{golov1,golov2,debenedictis,SSpaper,TdSpaper,nonvacSSpaper,
nonvacKSpaper,scalarfieldSSpaper,scalarfieldKSpaper,roberthudsonSSpaper,baha1,
bahagolov1,awad1,baha6,nashed5,pfeifer2,elhanafy1,benedictis3,baha10,baha4,
ruggiero1,ruggiero2,sahoo1,sahoo2,calza,sharif2009teleparallel,
pfeifer2022quick,attractor}. In $F(T)$ gravity, 
exact spherically symmetric configurations have been obtained for vacuum, 
scalar-field, and perfect-fluid sources, as well as for various anisotropic 
fluids related to dark-energy models 
\cite{Gonzalez:2011dr,Ferraro:2011ks,Capozziello:2012zj,Nashed:2013bfa,
Nashed:2013owg,Mai:2017riq,Golovnev:2021lki,Nashed:2021pah,Ren:2021uqb,
Papanikolaou:2022hkg,Awad:2022fhx,Zhao:2022gxl,DeBenedictis:2022sja,
Wang:2023qfm,Addazi:2023pfx,Wu:2024vcr,Rani:2022tmf,Jamil:2012ti,
landryvandenhoogen1,bahamonde2021exploring,salvatore1,scalarfieldTRW,
coleylandrygholami,FTBcosmogholamilandry}. 
These solutions have demonstrated that 
torsional corrections can effectively play the role of exotic matter, leading 
to traversable wormhole geometries without explicit violation of energy 
conditions. On the other hand, massive gravity theories have also provided new 
classes of 
static or dynamical wormholes, in which the massive term contributes an 
effective anisotropic stress that supports the throat structure. Nevertheless, 
the combination of $F(T)$ gravity with a weak dRGT massive term remains largely 
unexplored, and its potential to generate physically acceptable wormhole 
solutions has not been fully examined.

In the present work, we construct and analyze static, spherically symmetric 
traversable wormhole configurations in massive $F(T)$ gravity, where the 
standard teleparallel action is supplemented by a weak dRGT-type massive term. 
We work using orthonormal frames
where all the extra degrees of freedom are well-defined, and we   solve 
the antisymmetric parts of the FEs for 
the spin-connection components before solving their symmetric parts, using 
coframe component 
ansätze \cite{SSpaper,scalarfieldSSpaper,nonvacSSpaper,nonvacKSpaper,TdSpaper,
scalarfieldKSpaper,roberthudsonSSpaper,bahabohmer}. { The orthonormal frames, in comparison with the proper frames, have the advantage of well-defined degrees of freedom for the FE components. The proper frames have a main weakness that the antisymmetric components are all trivial, leaving a number of extra degrees of freedom. This last situation can potentially lead to some valid solutions on a proper frame, and then the same ones become invalid by only changing to a non-proper frame. To avoid this    we work in the current paper on an orthonormal non-proper frame.}

In particular, using the Morris-Thorne metric ansatz and appropriate 
orthonormal tetrads, we derive the full set of field and conservation equations (eqns), 
separating the effective energy-momentum tensor into cosmological-fluid and 
massive-graviton components. We then consider several redshift functions 
(constant, logarithmic, and power-law forms) and reconstruct the corresponding 
shape functions and $F(T)$ models. For each case, we examine the flaring-out 
condition and the fulfillment or controlled violation of the energy conditions (ECs),
since it is known that the extra torsion terms can   produce effective 
anisotropic pressures 
that can mimic familiar matter at large radii, while still preserving the 
essential throat geometry near $r = 
r_0$ 
\cite{tamanini2012tetrads,bamba2011singularity,sharif2013wormholes,
davis2015anisotropic}.
The obtained results demonstrate that massive teleparallel gravity admits 
self-consistent wormhole solutions which remain asymptotically flat and do not 
require the introduction of additional exotic sources.  In addition, we   
take into account the dark energy (DE) and/or any cosmological fluid, since  
these additional   sources can lead to new    solutions. Some other wormholes models in modified gravity can be seen in following Refs.~\cite{Shweta:2023hvm, Gudekli:2022vjs,Hussain:2022lxb,Abbas:2023jhi,Battista:2024gud,DeFalco:2021ksd,DeFalco:2021klh,Zubair:2023dbe,Ahmed:2022myl,Ashraf:2023bfg,Naz:2022smd}

The plan of the manuscript is the following. In Section \ref{sect2} we present 
the formalism and the eqns in massive $F(T)$ gravity, and in particular the 
 tetrad and spin-connection pair eqns, the FEs, the conservation laws
(CLs) and the ECs for the wormhole ansatz. 
Additionally, we consider three cases for the metric redshift function, namely 
the constant, the logarithmic, and the power-law ones. Then in sections 
\ref{sect4}, \ref{sect5}, and \ref{sect6} we extract exact solutions for the 
aforementioned three metric redshift function forms, for both the general 
massive gravity case as well as the massive gravity with uniform pressure case. 
In section \ref{sectphys} we analyze the physical features of the results, and 
we will perform a comparison between the various cases. Finally,  \ref{sect6}  is 
devoted to the conclusions.

\section{Spherically symmetric solutions in massive $F(T)$ gravity}\label{sect2}

In this section we briefly review massive { teleparallel} $F(T)$ gravity and we present the 
spherically symmetric solutions.

\subsection{Massive { teleparallel} $F(T)$ gravity and field equations}

The massive  $F(T)$ gravity  action     is 
\cite{Aldrovandi_Pereira2013,Bahamonde:2021gfp,Krssak:2018ywd,Coley:2019zld,
SSpaper,nonvacSSpaper,scalarfieldSSpaper,roberthudsonSSpaper,TdSpaper}:
\begin{equation}\label{1000}
	S=\int d^4x\,\left\{\frac{h}{2\kappa}\,F(T)+\mathcal{L}_{\rm matter}+\mathcal{L}_{\rm mass}\right\},
\end{equation}
with $h=\det(h^a{}_\mu)$ is the coframe determinant, and where $\mathcal{L}_{\rm 
matter}$ is the usual matter Lagrangian and $\mathcal{L}_{\rm mass}$ 
is the massive Lagrangian    for the 4-dimensional spacetime.
 The torsion tensor $T^a_{~~\mu\nu}$, the torsion 
scalar $T$ and the super-potential $S_a^{~~\mu\nu}$ are defined as 
\cite{Coley:2019zld}:
\begin{align}
	T^a_{~~\mu\nu} =& 
\partial_{\mu}\,h^a_{~~\nu}-\partial_{\nu}\,h^a_{~~\mu}+\omega^a_{~~b\mu}h^b_{
~~\nu}-\omega^a_{~~b\nu}h^b_{~~\mu}, \label{torsionten}
	\\
	S_a^{~~\mu\nu}=& 
\frac{1}{2}\,\left(T_a^{~~\mu\nu}+T^{\nu\mu}_{~~a}-T^{\mu\nu}_{~~a}\right)-h_a^{
~~\nu}\,T^{\lambda\mu}_{~~\lambda}+h_a^{~~\mu}\,T^{\lambda\nu}_{~~\lambda},
	\\
	T=&\frac{1}{2}\,T^a_{~~\mu\nu}S_a^{~~\mu\nu}. \label{torsionscal}
\end{align}	
%Eqn. \eqref{torsionten} can be expressed in terms of the three irreducible 
%parts 
%of torsion tensor as:
%\begin{align}
%    T_{abc} = 
%\frac{2}{3}\left(t_{abc}-t_{acb}\right)-\frac{1}{3}\left(g_{ab}V_c-g_{ac}
%V_b\right)+\epsilon_{abcd}A^d
%\end{align}
%where,
%\begin{align}
%    V_a=&T^b_{~ba} ,\quad A^a=\frac{1}{6}\epsilon^{abcd}T_{bcd} ,\quad
%    t_{abc}= 
%\frac{1}{2}\left(T_{abc}+T_{bac}\right)-\frac{1}{6}\left(g_{ca}V_b+g_{cb}
%V_a\right)+\frac{1}{3}V_c.
%\end{align}
Additionally, $\mathcal{L}_{\rm mass}$  is given by
\cite{massivefrgravity1,derham1,derham2,derham3,derham4,derham5,derham6}:
\begin{align}\label{masslagrangian}
	\mathcal{L}_{mass} =& \frac{m^2}{4}\sum_{i=1}^{4} c_i\,\mathcal{U}_i
\end{align}
where $m$ is the graviton mass, $c_i$ are constant parameters, and the 
$\mathcal{U}_i$ are the elementary symmetric polynomials of the eigenvalues of 
$K^\mu{}_\nu=\delta^\mu{}_\nu-\sqrt{g^{\mu\alpha}f_{\alpha\nu}}$
with $f_{\mu\nu}$ the Minkowski reference metric.

Varying action \eqref{1000}  we obtain the symmetric and 
antisymmetric parts of FEs, namely \cite{SSpaper} 
\begin{eqnarray}
		\kappa\,\Theta_{\left(ab\right)} &=& F_T \overset{\ 
\circ}{G}_{ab}+F_{TT}\,S_{\left(ab\right)}^{\;\;\;\mu}\,\partial_{\mu} 
T+\frac{g_{ab}}{2}\,\left[F-T\,F_T\right],  \label{1001a}
		\\
		0 &=& F_{TT}\,S_{\left[ab\right]}^{\;\;\;\mu}\,\partial_{\mu} T, 
\label{1001b}
	\end{eqnarray}
where $\overset{\ \circ}{G}_{ab}$ is the Einstein tensor, 
$\Theta_{\left(ab\right)}$ the energy-momentum, $g_{ab}$ the gauge metric and 
$\kappa$ the coupling constant.
We mention that the above  energy-momentum arises as
\begin{align}\label{1001ca}
	\Theta_a^{\;\;\mu}=\frac{1}{h} \frac{\delta}{\delta 
h^a_{\;\;\mu}}\left(\mathcal{L}_{\rm matter}+\mathcal{L}_{\rm 
mass}\right)\equiv \left[\Theta_{(ab)}\right]_{\rm 
mass}+\left[\Theta_{(ab)}\right]_{\rm matter},
\end{align}
with $\left[\Theta_{(ab)}\right]_{\rm matter}$   the usual matter 
energy-momentum tensor, and where the effective
  massive, dRGT energy-momentum after variation acquires the form
\cite{massivefrgravity1,derham1,derham2,derham3,derham4,derham5,derham6}:
\begin{align}\label{massem}
	\left[\Theta_{(ab)}\right]_{\rm mass} = 
\frac{m^2}{2}\left[\left(c_1\mathcal{U}_1+c_2\mathcal{U}_2\right)g_{ab}
-\left(c_1+2c_2\mathcal{U}_1\right)\left[\mathcal{K}_{ab}\right]+2c_2\left[
\mathcal{K}_{ab}^2\right]\right],
\end{align}
with $\mathcal{U}_0=1$, $\mathcal{U}_1=\left[\mathcal{K}\right]^a_{~a}$, and 
$\mathcal{U}_2=\left(\left[\mathcal{K}\right]^a_{~a}\right)^2-\left[\mathcal{K}
^2\right]^a_{~a}$. 
As we see from   \eqref{massem}, there are two typical types of 
massive structures, and in the following we  assume that   
\eqref{massem} describes a weak dRGT model with negligible coupling effects in 
comparison to geometry, weak massive graviton, and cosmological fluid separated 
terms. This approach allows us to simplify the     analysis   by keeping the 
same orthonormal frame and FE system.

\subsection{General static spherically symmetric tetrad/spin-connection pairs}

In~the orthonormal gauge $g_{ab} =\eta_{ab}= Diag[-1,1,1,1]$, any teleparallel geometry satisfies the relations~\cite{olver1995equivalence}:
%\begin{subequations}
\begin{align}
	\mathcal{L}_{{\bf X}} {\bf h}^a =& \lambda^a_{~b} {\bf h}^b \quad\text{and}\quad  \mathcal{L}_{{\bf X}} {\omega}^a_{~bc} = 0, \label{Intro12}
\end{align}
%\end{subequations}
where ${\bf h}^a$ is the  {differential form of} the tetrad,  
$\mathcal{L}_{{\bf X}}$ is the Lie derivative in terms of Killing Vectors (KV) 
${\bf X}$, and $\lambda^a_{~b}$ is the generator of  $\Lambda^a_{~b}$. Moreover, 
  for a pure teleparallel geometry one must satisfy the zero-curvature 
requirement, namely   
\cite{Lucas_Obukhov_Pereira2009,Aldrovandi_Pereira2013,Bahamonde:2021gfp, 
MCH,Krssak:2018ywd,Coley:2019zld,Krssak_Pereira2015}:
\begin{align}\label{zerocurvatureeqsol}
	R^a_{~b\mu\nu}  =& \partial_{\mu}\omega^a_{~b\nu} -\partial_{\nu}\omega^a_{~b\mu}+\omega^a_{~e\mu}\omega^e_{~b\nu}-\omega^a_{~e\nu}\omega^e_{~b\mu} =0 ,
	\nonumber\\
	\Rightarrow\quad &\omega^a_{~b\mu} = 
\Lambda^a_{~c}\partial_{\mu}\Lambda_b^{~c}, 
\end{align} 
and the solution of   \eqref{zerocurvatureeqsol} leads to the teleparallel 
spin-connection defined in terms of $\Lambda^a_{~b}$. Note  that 
$\omega^a_{~b\mu}=0$ for all proper frames and   $\omega^a_{~b\mu}\neq 0$ for 
all non-proper frames.

%Add justifications on AFEs using leading to coframe/spin-connection pair.

The teleparallel spherically symmetric spacetimes were     discussed in 
detail   in  \cite{SSpaper}. As the authors showed, 
there is a { coframe}/spin-connection pair ``diagonalizing'' the frame described 
by a static spherically symmetric and orthonormal vierbein, namely 
\cite{SSpaper,nonvacSSpaper} 
\begin{align}
	& h^a_{~\mu} =  \text{Diag}\left[ A_1(r),\, A_2(r),\,A_3(r),\,A_3(r) \sin(\theta)\right], \label{1100}
\\
& \omega_{233} = \omega_{244} =  \frac{\delta}{A_3},~ \omega_{344} = - \frac{\cos(\theta)}{A_3 \sin(\theta)},\label{1101}
\end{align}
where $\delta=\pm 1$. { The $\delta$-parameter value in   \eqref{1101} will determine the type of inertial effects for a typical spherically symmetric spacetime. Relations \eqref{1101} describing the spin-connection components are obtained by solving the antisymmetric parts of the field equations as developed and justified in detail in ref. \cite{SSpaper}.} The pair of coframe/spin connections   
\eqref{1100}-\eqref{1101} goes in the same direction and improves the 
expressions obtained recently in refs \cite{Krssak:2018ywd,golov3}. 

Let us now proceed to the wormhole solutions. As it is known a wormhole 
metric can be defined under the Morris-Thorne ansatz as
\cite{massivefrgravity1,morristhorne1,morristhorne2} 
\begin{align}\label{1104}
	ds^2 = - \exp\left[2\Phi(r)\right]dt^2+ 
\left[1-\frac{b(r)}{r}\right]^{-1}\,dr^2 + r^2d\theta^2 + 
r^2\,\sin^2\theta\,d\phi^2,
\end{align}
where $\Phi(r)$ is the redshift parameter and $b(r)$  is a horizon ansatz 
parameter, also called   the shape function. Hence, under this ansatz,
namely under $A_1=\exp\left(\Phi(r)\right)$, 
$A_2=\frac{1}{\sqrt{1-\frac{b(r)}{r}}}$ and $A_3=r$, 
the 
eqn pair \eqref{1100}-\eqref{1101}  becomes 
\begin{align}
& h^a_{~\mu} = \text{Diag}\left[ \exp\left(\Phi(r)\right),\, {\left(1-\frac{b(r)}{r}\right)}^{-1/2},\,r,\,r \sin(\theta)\right] , \label{1102}
\\
& \omega_{233} = \omega_{244} = \frac{\delta}{r},~ \omega_{344} = - \frac{\cos(\theta)}{r\,\sin(\theta)}.  \label{1103}
\end{align}
We mention here  that in   \eqref{1101} and \eqref{1103} the spin-connection 
depend only of the $A_3$ component, which implies that the $A_1$ and 
$A_2$ components in  \eqref{1104} will not affect the spin connection and 
will only affect the { coframe components as also shown in  \cite{SSpaper}}.

\subsection{General fluid environment and  energy 
conditions}

Let us now analyze the matter environment of the present analysis.
As it is known, the perfect fluid energy-momentum   is 
\cite{roberthudsonSSpaper} 
\begin{align}\label{PFEM}
T_{ab}=(\rho+P_t)u_a\,u_b+g_{ab}\,P_t+\left(P_r-P_t\right)v_av_b ,
\end{align}
where $u_a u^a=-1$, $v_av^a=+1$,  $P_r$ is the radial pressure and $P_t$ is the 
tangential pressure. In the  scenario studied in the present work, we  can 
split the total energy-momentum tensor   as
\begin{align}\label{weakmassiveEM}
T_{ab}=T^{CF}_{ab}+T^{(\rm mass)}_{ab},
\end{align}
where $T^{CF}_{ab}$ the perfect cosmological fluid energy-momentum and  $T^{(\rm 
mass)}_{ab}$ the ``weak'' massive   energy-momentum.  We consider a   
cosmological perfect fluid with equation of state (EoS)   
$P_{CF}=\alpha_{CF}\,\rho_{CF}$ (note that $P_r=P_t$ for the cosmological 
fluid). { For a more realistic     scenario, we will assume a dark energy-dominated universe background,
as it was done in recent works on teleparallel $F(T)$ gravity   \cite{nonvacSSpaper,nonvacKSpaper,roberthudsonSSpaper,scalarfieldSSpaper,scalarfieldKSpaper}. This last background assumption will easily allow to compare the massive source solutions to those of a typical dark energy-dominated universe by using the same orthonormal coframe/spin-connection pair and then the same form of field equations.} Additionally, we assume a weak massive graviton model, allowing us to 
use \eqref{PFEM}-\eqref{weakmassiveEM}, and then allowing us to neglect the 
coupling between gravitons and the wormhole geometry for a simpler analysis 
using orthonormal gauge.

In summary, the total energy density and pressure are
\begin{align}
\rho=\rho_{CF}+\rho_{\rm mass},\quad P_r=\alpha_{CF}\,\rho_{CF}+{P_r}_{\rm mass}, \quad P_t=\alpha_{CF}\,\rho_{CF}+{P_t}_{\rm mass} . \label{1300}
\end{align}
Furthermore, the CLs in terms of tetrad components, for $A_3=r$, 
are written  as \cite{roberthudsonSSpaper} 
\begin{align}
    \partial_rP_r+\left(P_r+\rho\right) \partial_r\left(\ln\, 
A_1\right)+\frac{2}{r}\left(P_r-P_t\right)  =0. \label{1301} 
\end{align} 
One could further restrict to the case  $A_3=c_0=const.$ 
and thus   \eqref{1301} becomes
\begin{align}
    \partial_rP_r+\left(P_r+\rho\right)\partial_r\left(\ln\, A_1\right)=0, 
\label{1301special}
\end{align}
eqn that is independent from the tangential pressure $P_t$.

We can now write the general FEs   of 
massive $F(T)$ gravity with a cosmological fluid, using relations 
\eqref{1100}-\eqref{1101} and with $A_3=r$, as
\cite{SSpaper,nonvacSSpaper,scalarfieldSSpaper,roberthudsonSSpaper}
\begin{align}
\kappa \left(\rho_{CF}+\rho_{\rm mass}\right)=& 
-\frac{1}{2}\left[F-TF_T\right]-2\partial_r\left(F_T\right)\,\left[\frac{\delta}
{rA_2}+\frac{1}{rA_2^2}\right]+F_T\left[\frac{2A_2'}{rA_2^3}-\frac{1}{r^2A_2^2}
+\frac{1}{r^2}\right] ,\label{1302}
    \\
\kappa \left(\alpha_{CF}\rho_{CF}+{P_r}_{\rm mass}\right)=& 
\frac{1}{2}\left[F-TF_T\right]+F_T\left[\frac{2A_1'}{rA_1A_2^2}+\frac{1}{
r^2A_2^2}-\frac{1}{r^2}\right] , \label{1303}
    \\
\kappa \left(\alpha_{CF}\rho_{CF}+{P_t}_{\rm mass}\right)=& 
\frac{1}{2}\left[F-TF_T\right]+\partial_r\left(F_T\right)\,\left[\frac{\delta}{
rA_2}+\frac{A_1'}{A_1 A_2^2}+\frac{1}{rA_2^2}\right]
+F_T\Bigg[\frac{A_1''}{A_1A_2^2}-\frac{A_1'A_2'}{A_1A_2^3}+\frac{A_1'}{rA_1A_2^2
}-\frac{A_2'}{rA_2^3}\Bigg] , \label{1304}
\end{align}
with
 \begin{align}
 T(r) =&  
-2\left(\frac{\delta}{r}+\frac{1}{rA_2}\right)\left(\frac{\delta}{r}+\frac{1}{
rA_2}+\frac{2\,A_1'}{A_1\,A_2}\right).    \label{1305}
\end{align}

Finally, we can examine the validity of various ECs, such as the
\cite{Kontou:2020bta}:
\begin{itemize}
    \item Weak EC (\textbf{WEC}): $\rho \geq 0$, $P_r+\rho \geq 0$ and $P_t+\rho \geq 0$.

    \item Strong EC (\textbf{SEC}): $P_r+2P_t+\rho \geq 0$, $P_r+\rho \geq 0$ and $P_t+\rho \geq 0$.

    \item Null EC (\textbf{NEC}): $P_r+\rho \geq 0$ and $P_t+\rho \geq 0$. 

    \item Dominant EC (\textbf{DEC}): $\rho \geq |P_r|$ and $\rho \geq |P_t|$.
\end{itemize}
The above ECs can be written collectively as:
 \begin{align}
& \rho+P_r \geq 0 ,  \quad \rho+P_t \geq 0 , \quad \rho+P_r+2P_t\geq 0 , \quad 
\rho \geq |P_r|,  \quad \rho \geq |P_t|, \quad\rho \geq 0.   \label{1409}
\end{align}

In summary, eqns \eqref{1301}-\eqref{1305} are the basic eqns used in 
the present manuscript. One can use any 
$A_1$-$A_2$ coframe ansatz and any   source as EoS parameter, and then 
  find the solution for $F(T)$ form
and  any other physical quantity. eqns 
\eqref{1302}-\eqref{1305} are relevant for a weak massive graviton scenario 
that  allows us to neglect the possible geometry-mass coupling effects.

\subsection{General massive gravity  case}
\label{gencase}

Let us now focus on the massive gravity sector. Following 
\cite{massivefrgravity1} we set  
$\mathcal{K}_{ab}=\text{Diag}\left[0,\,0,\,K_3(r),\,K_3(r)\right]$,     finding
that $\mathcal{U}_1=2K_3(r)$, $\mathcal{U}_2=2K_3(r)^2$. In this case, relation 
\eqref{massem} becomes 
\begin{align}\label{massemc1}
	\left[\Theta_{(ab)}\right]_{\rm mass} =&  
\frac{m^2}{2}\left[\left(2c_1K_3(r)+2c_2K_3(r)^2\right)g_{ab}
-\left(c_1+4c_2K_3(r)\right)\left[\mathcal{K}_{ab}\right]+2c_2\left[\mathcal{K}_
{ab}^2\right]\right],
\end{align}
where
$	\rho_{\rm mass}={P_r}_{\rm mass}= m^2K_3\left(c_1+c_2\,K_3\right)$ and
	${P_t}_{\rm mass}= \frac{c_1\,m^2}{2}\,K_3,$
with $K_3=K_3(r)$. Moreover, the CL  
\eqref{1301} reads  
\begin{align}
	&\alpha_{CF} \partial_r\rho_{CF}+\left(1+\alpha_{CF}\right)\rho_{CF}\Phi'+m^2\Bigg[\left(\partial_rK_3+\frac{K_3}{r}\right)\left(c_1+2c_2\,K_3\right)+K_3\left(c_1+c_2\,K_3\right)\Phi'\Bigg] =0. \label{1401}
\end{align}
Finally, the ECs  \eqref{1409} become
\begin{align}
	\left(1+\alpha_{CF}\right)\rho_{CF}+2m^2K_3\left(c_1+c_2\,K_3\right) \geq 0 , \label{1406b}
	\\
	\left(1+\alpha_{CF}\right)\rho_{CF}+\frac{m^2K_3}{2}\left(3c_1+2c_2\,K_3\right) \geq 0,  \label{1408b}
	\\
	\left(1+3\alpha_{CF}\right)\rho_{CF}+m^2K_3\left(3c_1+2c_2\,K_3\right) \geq 0 , \label{1407b}
	\\
	\rho_{CF}+m^2K_3\left(c_1+c_2\,K_3\right) \geq |\alpha_{CF}\rho_{CF}+m^2K_3\left(c_1+c_2\,K_3\right)|, \label{1409b}
	\\
	\rho_{CF}+m^2K_3\left(c_1+c_2\,K_3\right) \geq |\alpha_{CF}\rho_{CF}+\frac{c_1\,m^2}{2}\,K_3|, \label{1409c}
	\\
	\rho_{CF}+ m^2K_3\left(c_1+c_2\,K_3\right)\geq 0. \label{1409d}
\end{align}

{ Under the null graviton mass $m\,\rightarrow\,0$ limit, Eqn. \eqref{massemc1} vanishes, Eqn. \eqref{1401} will only keep the cosmological fluid terms, and Eqns. \eqref{1406b}--\eqref{1409d} will be greatly simplified by only keeping the cosmological dark energy fluid terms as  it was done in \cite{nonvacSSpaper,scalarfieldSSpaper,roberthudsonSSpaper}. We will find that $-1\leq \alpha_{CF} \leq 1$ and $\rho_{CF} \geq 0$ for most of the energy conditions, and $\left(1+3\alpha_{CF}\right)\rho_{CF}\geq 0$ from   \eqref{1407b}. Relation \eqref{1407b} can be considered as questionable, since it may leave a doubt on dark energy existence itself. For the $m\neq 0$ case,   \eqref{1407b} makes a direct relation between massive solitons and dark energy.} Let us now extract analytical solutions of the above eqns. These may 
be possible in the following cases:
\begin{enumerate}
	\item \textbf{Constant} $\Phi(r)=\Phi_0$: By setting $\Phi'(r)=0$ in   
\eqref{1401} and assuming that the cosmological fluid does not directly interact 
with the massive gravitons, the separate CLs are 
	\begin{align}
		0=& \alpha_{CF} \partial_r\rho_{CF} , \label{1411}
		\\
		0=& \left(\partial_rK_3+\frac{K_3}{r}\right)\left(c_1+2c_2\,K_3\right) . \label{1412}
	\end{align}
Eqn \eqref{1411} yields to  
	 $\rho_{CF}=const.$   and $\alpha_{CF} \neq 0$, or to $\alpha_{CF}=0$ (dust 
fluid). 
	Additionally, eqn \eqref{1412} for a non-constant $K_3$ yields to the 
same solution  obtained in \cite{massivefrgravity1}, namely
	\begin{align}\label{Newtonianmassive}
		K_3(r)=\frac{C}{r}, 
	\end{align}
	where $C$ is an integration constant.

	\item \textbf{Logarithmic} $\Phi(r)=\Phi_0+a\,\ln (r)$: In this case 
eqn \eqref{1401}  gives rises to
	\begin{align}
		0=& \alpha_{CF} \partial_r\rho_{CF}+\left(1+\alpha_{CF}\right)\frac{a}{r}\rho_{CF} , \label{1421}
		\\
		0=& \left(\partial_rK_3+\frac{K_3}{r}\right)\left(c_1+2c_2\,K_3\right)+K_3\left(c_1+c_2\,K_3\right)\frac{a}{r} . \label{1422}
	\end{align}
	Eqn \eqref{1421} for $\alpha_{CF}\neq 0$ leads to a power-law fluid 
density \cite{nonvacSSpaper}:
	\begin{align}\label{1423}
		\rho_{CF}(r)=\rho_{CF}(0)\,r^{-\frac{a\left(1+ 
\alpha_{CF}\right)}{\alpha_{CF}}},
	\end{align} 
	while for $\alpha_{CF}=0$  we find that $a=0$ and then we recover the 
$\Phi=\Phi_0=$ constant redshift case. Then,  \eqref{1422} will lead 
to the solution 
	\begin{align}\label{1424}
	\frac{C}{r}=\left[(1+a)c_1+(2+a)c_2\,K_3\right]^{\frac{a}{(1+a)(2+a)}}\,K_3^{\frac{1}{(1+a)}},   
	\end{align}
	where $C$ is the integration constant. Relation \eqref{1424} is an implicit 
form of solution where an explicit form of $K_3(r)$ will depend on the value of 
$a$. There are   two limiting cases:
	\begin{itemize}
		\item $c_1\ll c_2$ and/or $a=-1$: Setting $a=-1+\Delta a$ where $\Delta 
a \ll 1$,   \eqref{1424} simplifies to:
		\begin{align}\label{1425}
			K_3(r)\approx \frac{\tilde{C}}{\sqrt{r}} .
		\end{align}
				\item $c_1\gg c_2$ and/or $a=-2$: Setting $a=-2+\Delta a$,   
\eqref{1424} simplifies   to:
		\begin{align}\label{1426}
			K_3(r) \approx \frac{r}{\tilde{C}}  ,
		\end{align}
	\end{itemize}
	where $\tilde{C}$ is the modified integration constant.

	\item \textbf{Power-law} $\Phi(r)=\Phi_0+a_1\,r^a$: In this case eqn 
\eqref{1401}  gives rises to
	\begin{align}
		0=& \alpha_{CF} \partial_r\rho_{CF}+\left(1+\alpha_{CF}\right)a_1 a r^{a-1}\rho_{CF} , \label{1431}
		\\
		0=& \left(\partial_rK_3+\frac{K_3}{r}\right)\left(c_1+2c_2\,K_3\right)+K_3\left(c_1+c_2\,K_3\right) a_1 \left(r^{a}\right)' . \label{1432}
	\end{align}
	The solution of   \eqref{1431} is 
	\begin{align}\label{1433}
		\rho_{CF}(r) = \rho_{CF}(0)\,\exp\left(-\frac{\left(1+ 
\alpha_{CF}\right)}{\alpha_{CF}}\,a_1\,r^a\right),
	\end{align}
	while  \eqref{1432} can be solved only by assuming 
$K_3(r)=\frac{g(r)}{r}$, leading to 
	\begin{align}\label{1434}
		-a_1\,r^a=\int_{r} 
dr'\,\frac{\partial_{r'}\,g(r')\left(c_1+2c_2\,\frac{g(r)}{r}\right)}{
g(r')\left(c_1+c_2\,\frac{g(r)}{r}\right)},
	\end{align}
	which has two limiting cases:
	\begin{itemize}
		\item $c_2\ll c_1$, which leads to 
		\begin{align}\label{1435}
			& K_3(r)=g(0)\,\frac{e^{-a_1\,r^a}}{r}.
		\end{align}
		
		\item $c_2\gg c_1$, which leads to 
		\begin{align}\label{1436}
			& K_3(r)=g(0)\,\frac{  e^{-\frac{a_1}{2}\,r^a}}{r} .
		\end{align}
			\end{itemize}
	As we see,  \eqref{1435}-\eqref{1436}  describe a Yukawa-like 
type of solutions.  
	
\end{enumerate}

In summary, inserting all the above into  \eqref{1302}-\eqref{1305} yields the 
eqns
\begin{align}
	\kappa \left(\rho_{CF}+m^2K_3\left(c_1+c_2\,K_3\right)\right)
	=& -\frac{1}{2}\left[F-TF_T\right]-2\partial_r\left(F_{T}\right)\,\frac{\left(1-\frac{b(r)}{r}\right)^{1/2}}{r}\left[\delta+\left(1-\frac{b(r)}{r}\right)^{1/2}\right]+{F_T}\,\frac{b'(r)}{r^2} ,\label{1402}
	\\
	\kappa \left(\alpha_{CF}\rho_{CF}+m^2K_3\left(c_1+c_2\,K_3\right)\right)=& \frac{1}{2}\left[F-TF_T\right]+F_T\left[\frac{2r\Phi'}{r^2}-\frac{b(r)}{r^3}\left(2r\Phi'+1\right)\right] , \label{1403}
	\\
	\kappa \left(\alpha_{CF}\rho_{CF}+\frac{c_1\,m^2}{2}\,K_3\right)=& \frac{1}{2}\left[F-TF_T\right]+\partial_r\left(F_{T}\right)\,\,\frac{\left(1-\frac{b(r)}{r}\right)^{1/2}}{r}\left[\delta+\left(r \Phi'+1\right)\left(1-\frac{b(r)}{r}\right)^{1/2}\right]
	\nonumber\\
	&\,+F_T\Bigg[\left(\Phi'^2+\Phi''\right)\left(1-\frac{b(r)}{r}\right)+\frac{2r^2\Phi'+b(r)(1-r\Phi')-rb'(r)(r\Phi'+1)}{2r^3}\Bigg] , \label{1404}
	\\
	T(r) =&  -\frac{2}{r^2}\left(\delta+\left(1-\frac{b(r)}{r}\right)^{1/2}\right)\left(\delta+\left(1+2\,r\Phi'\right)\left(1-\frac{b(r)}{r}\right)^{1/2}\right). \label{1405}
\end{align}
Eqns \eqref{1402}-\eqref{1405} are the FEs for general massive $F(T)$ gravity in terms of $\Phi(r)$, $b(r)$ and $K_3(r)$, 
within a cosmological fluid. In the following sections we will solve  them by 
using   specific ansätze for $r(T)$, $b(r(T))$ and $K_3(r(T))$. We will also 
study   specific  $F(T)$ functions, and we will find possible $b(r(T))$ and 
$K_3(r(T))$. { The redshift choices are justified by specific and typical physical properties. The constant redshift case will essentially lead to static cosmological-like teleparallel $F(T)$ solutions, allowing one to easily compare solutions with Teleparallel Robertson-Walker (TRW) and Kantowski-Sachs (KS) teleparallel $F(T)$ solutions \cite{nonvacKSpaper,scalarfieldKSpaper}. The logarithmic reshift case is exactly the pure power-law coframe ansatz as used in most of the refs \cite{SSpaper,nonvacSSpaper,scalarfieldSSpaper} and will allow us to compare the massive solutions with the pure dark energy teleparallel $F(T)$ solutions found in recent papers \cite{nonvacSSpaper,scalarfieldSSpaper}. Finally, the power-law redshift or the exponential coframe ansatz can be interpreted as a generalization of logarithmic redshift (or power-law coframe) case, where an exponential is fundamentally an infinite sum of power-law terms defined as $\exp\left(\Phi(r)\right)=\sum_{p\geq0}\,\frac{\Phi^p}{p!}$. The redshift choices reflect the possible physical systems and cases leading to typical and consistent solutions by highlighting the weak massive source contributions to the new teleparallel $F(T)$ solutions.}

\subsection{Massive gravity with uniform pressure}

 Let us now impose a different ansatz, and set 
$\mathcal{K}_{ab}=\text{Diag}\left[0,\,K_3(r),\,K_3(r),\,K_3(r)\right]$. In 
this case we find that $\mathcal{U}_1=3K_3(r)$, and $\mathcal{U}_2=6K_3(r)^2$. 
Hence, eqn  \eqref{massem} becomes 
\begin{align}\label{massem2}
	\left[\Theta_{(ab)}\right]_{\rm mass}  =& 
\frac{m^2}{2}\left[\left(3c_1K_3(r)+6c_2K_3(r)^2\right)g_{ab}
-\left(c_1+6c_2K_3(r)\right)\left[\mathcal{K}_{ab}\right]+2c_2\left[\mathcal{K}_
{ab}^2\right]\right] ,
\end{align}
where
\begin{align}
	\rho_{mass}=& \frac{3m^2}{2}\,K_3\left(c_1+2c_2K_3\right),
	\quad {P_r}_{mass}={P_t}_{mass}=P_{mass}= m^2K_3\left(c_1+c_2K_3\right).
\end{align}
Additionally, eqn \eqref{1301} reads
\begin{align}
	& \alpha_{CF} \partial_r\rho_{CF}+\left(1+\alpha_{CF}\right)\rho_{CF}\Phi' +m^2\Bigg[\partial_rK_3\left(c_1+2c_2K_3\right)+\frac{K_3}{2}\left(5c_1+8c_2K_3\right)\Phi'\Bigg]  =0. \label{1501}
\end{align}
We introduce the  EoS parameter $\alpha_{mass}$  as
\begin{align}
	\alpha_{mass} 
\equiv \frac{P_{mass}}{\rho_{mass}}=\frac{2\left[c_1+c_2K_3(r)\right]}{
3\left[c_1+2c_2K_3(r)\right]}.
\end{align} 
Finally, the ECs \eqref{1409} become
\begin{align}
	\left(1+\alpha_{CF}\right)\rho_{CF}+\frac{m^2}{2}\,K_3\left(5c_1+8c_2K_3\right) \geq 0 , \label{1506}
	\\
	\left(1+3\alpha_{CF}\right)\rho_{CF}+\frac{3m^2}{2}\,K_3\left(3c_1+4c_2K_3\right) \geq 0 , \label{1507}
	\\
	\rho_{CF}+\frac{3m^2}{2}\,K_3\left(c_1+2c_2K_3\right) \geq |\alpha_{CF}\rho_{CF}+m^2K_3\left(c_1+c_2K_3\right) |, \label{1508}
	\\
	\rho_{CF}+\frac{3m^2}{2}\,K_3\left(c_1+2c_2K_3\right) \geq 0 . \label{1509}
\end{align}

The possible $K_3(r)$ solutions to eqn \eqref{1501}  will   depend on the 
redshift parameter $\Phi(r)$. Similarly to the previous subsection { and using the same physical justifications}, there are 
 three cases.
 
\begin{enumerate}
	\item \textbf{Constant} $\Phi(r)=\Phi_0$: In this case \eqref{1501} 
gives \eqref{1411} and 
	$
		0=\partial_rK_3\left(c_1+2c_2K_3\right)$, with solution  $K_3=C=const.$, 
and then $\rho_{\rm mass}=m^2\rho_0$ and $P_{\rm mass} = m^2P_0$.

	\item \textbf{Logarithmic} $\Phi(r)=\Phi_0+a\,\ln (r)$:
	In this case  \eqref{1501} gives
\eqref{1421} and  
	\begin{align}\label{1521a}
		& 0=\partial_rK_3\left(c_1+2c_2K_3\right)+\frac{K_3}{2}\left(5c_1+8c_2K_3\right)\frac{a}{r},
			\end{align} which leads to 
		\begin{align}\label{1521}
 r^{-10a}=C\,\left(5c_1+8c_2K_3\right)\,K_3^{4},
		 		\end{align}
	where $C$ is an integration constant.  Eqn \eqref{1521} is an implicit 
function leading to a degree $5$ polynomial relation, with two limiting 
cases: $c_1\ll c_2$, which leads to $
			K_3(r) \approx \tilde{C}\,r^{-2a}.
	$, and  $c_1 \gg c_2$ which gives
$
			K_3(r) \approx \tilde{C}\,r^{-\frac{5a}{2}},$
 	with $\tilde{C}$   an integration constant.
	
	\item \textbf{Power-law} $\Phi(r)=\Phi_0+a_1\,r^a$: 
	In this case \eqref{1501}  gives  \eqref{1431} and
	\begin{align}\label{1531}
		&    
0=\partial_rK_3\left(c_1+2c_2K_3\right)+\frac{K_3}{2}
\left(5c_1+8c_2K_3\right)a_1\left(r^{a}\right)',
	\end{align}   
	 which leads to 
		\begin{align}\label{1531}
		 \exp\left(-10 
a_1\,r^{a}\right)=C\,\left(5c_1+8c_2K_3\right)\,K_3^{4},
	\end{align}  
	where $C$ is an integration constant.  Eqn \eqref{1531} is   an 
implicit function leading to a degree $5$ polynomial relation, with two 
limiting 
cases: $c_1 \ll c_2$ which yields
	$
			K_3(r) \approx \tilde{C}\,\exp \left(-2a_1\,r^a\right)$, and 
		  $c_1 \gg c_2$ which gives
	$
			K_3(r) \approx \tilde{C}\,\exp \left(-\frac{5a_1}{2}\,r^a\right),
$
	with $\tilde{C}$ is an integration constant.
\end{enumerate}

In summary, inserting all the above into  \eqref{1302}-\eqref{1305} yields the 
eqns 
\begin{align}
	\kappa \left(\rho_{CF}+\frac{3m^2}{2}\,K_3\left(c_1+2c_2K_3\right)\right)
	=& -\frac{1}{2}\left[F-TF_T\right]-2\partial_r\left(F_{T}\right)\,\frac{\left(1-\frac{b(r)}{r}\right)^{1/2}}{r}\left[\delta+\left(1-\frac{b(r)}{r}\right)^{1/2}\right]+{F_T}\,\frac{b'(r)}{r^2} ,\label{1502}
	\\
	\kappa \left(\alpha_{CF}\rho_{CF}+m^2K_3\left(c_1+c_2K_3\right)\right)=& \frac{1}{2}\left[F-TF_T\right]+F_T\left[\frac{2r\Phi'}{r^2}-\frac{b(r)}{r^3}\left(2r\Phi'+1\right)\right] , \label{1503}
	\\
	\kappa \left(\alpha_{CF}\rho_{CF}+m^2K_3\left(c_1+c_2K_3\right)\right)
	=&  \frac{1}{2}\left[F-TF_T\right]+\partial_r\left(F_{T}\right)\,\,\frac{\left(1-\frac{b(r)}{r}\right)^{1/2}}{r}\left[\delta+\left(r \Phi'+1\right)\left(1-\frac{b(r)}{r}\right)^{1/2}\right]
	\nonumber\\
	&\,+F_T\Bigg[\left(\Phi'^2+\Phi''\right)\left(1-\frac{b(r)}{r}\right)+\frac{2r^2\Phi'+b(r)(1-r\Phi')-rb'(r)(r\Phi'+1)}{2r^3}\Bigg] , \label{1504}
\end{align}
where the eqn for $T$ is again \eqref{1405}. Eqns \eqref{1405} and 
\eqref{1502}-\eqref{1504} are the FEs for uniformed pressure 
massive   $F(T)$ solutions in terms of $\Phi(r)$, $b(r)$ and 
$K_3(r)$. In the following sections we   impose  specific ansätze in terms of 
$r(T)$ and we express all previous physical quantities in terms of the torsion 
scalar $T$.

\section{Constant redshift function  solutions}\label{sect4}

In this section we examine the subclass of wormhole solutions that have a
constant redshift function. By setting  $\Phi(r)=\Phi_0=const.$ 
  (i.e. $A_1=a_0=const.$) and keeping $A_2$ under the same form, we find that eqn \eqref{1405} becomes
\begin{align}\label{2001}
	T(r) =  -\frac{2}{r^2}\left(\delta+\left(1-\frac{b(r)}{r}\right)^{1/2}\right)^2,
\end{align}
where $\delta=\pm 1$. { Equation \eqref{2001} will allow us to simplify the field equations for finding the new teleparallel $F(T)$ solutions as done in refs \cite{SSpaper,nonvacSSpaper,scalarfieldSSpaper}.} Similarly, the general characteristic eqn for   $b(r)$ is
	$0= \frac{T}{2} + 
\frac{\delta\,\delta_2}{r(T)}\sqrt{-2T}-\frac{b(r)}{r^3(T)}$
with $\delta_2=\pm 1$, and thus the shape function $b(T)$ in terms of the 
torsion scalar $T$ can be expressed as:
\begin{align}\label{2001shape}
	b(T)=\frac{T\,r^3(T)}{2}+\delta\,\delta_2\,\sqrt{-2T}r^2(T).
\end{align}

The above eqns   allow us to find $r(T)$ solutions by setting specific 
$b(r)$ functions. For instance, setting  $  
b(r)=2M+b_1\,r-\frac{\Lambda_0}{3}\,r^3 $ which for $M \ll \Lambda_0$ gives
	\begin{align}\label{2110m}
		 r^{-1}(T) \approx \frac{3 \sqrt{-2 T}\, \delta  \delta_{2}+\delta_3\sqrt{-18 T +18 b_{1} T+12 b_{1} \Lambda_{0}}}{6 b_{1}} ,
	\end{align}
	with $\delta_3=\pm 1$, and then $b(T)$ and $F(T)$ can be easily found.

\subsection{General massive gravity  case}

 Substituting eqns \eqref{Newtonianmassive} and \eqref{2001} into   
\eqref{1402}-\eqref{1404}, we obtain    the general   $F(T)$ solution 
as
\begin{align}\label{2108}
	F(T) = \kappa \left(1+3\alpha_{CF}\right)\rho_{0}+F_1\,T-\kappa\,m^2\,T\Bigg[3\tilde{c}_1\mathcal{F}_{-1}(T)+2\tilde{c}_2\mathcal{F}_{-2}(T)\Bigg] ,
\end{align}
where we have defined the   functions 
\begin{align}
	\mathcal{F}_{-1}(T)=\int_{T}\,\frac{dT'}{T'^2}\, r^{-1}(T'),\label{2109a}
	\\
	\mathcal{F}_{-2}(T)=\int_{T}\,\,\frac{dT'}{T'^2}\,r^{-2}(T').\label{2109b}
\end{align} 

We first impose the wormhole ansatz $ 
b(r)=2M+b_1\,r-\frac{\Lambda_0}{3}\,r^3$. In this case we find that  
	\small
	\begin{align}		
\mathcal{F}_{-1}(T)=&\frac{\delta_4\sqrt{3}\left(b_1-1\right)}{2\sqrt{b_{1} 
\Lambda_{0}} b_{1}} \mathrm{arctanh}\left(\frac{\sqrt{\left(18 T+12 
\Lambda_{0}\right) b_{1}-18 T}}{2 \sqrt{3b_{1} \Lambda_{0}}}\right)
		+\frac{1}{ b_{1}\,T} \left(\sqrt{-2T}\,\delta  
\delta_{2}+\frac{\delta_4}{6} \sqrt{\left(18 T+12 \Lambda_{0}\right) b_{1}-18 
T}\right) ,\label{2111a}
		\\
		\mathcal{F}_{-2}(T)=&  \frac{1}{b_{1}^{2}}\Bigg[\Bigg(\frac{2\,\delta_4 
\delta  \delta_{2}\sqrt{\left(3 T +2 \Lambda_{0}\right) b_{1}-3 
T}}{\sqrt{-3T}}+\left(\frac{b_{1}}{2}-1\right) \ln \left(T \right)-\frac{b_{1} 
\Lambda_{0}}{3T}\Bigg) 
		\nonumber\\
		&\,-\delta_4  \delta  \delta_{2} \sqrt{b_{1}-1} 
		\,\arctan \! \left(\frac{\left(\left(3T +\Lambda_{0}\right) b_{1}-3T 
\right) }{\sqrt{-3T}\sqrt{\left(3T +2\Lambda_{0}\right) b_{1}-3T}\, 
\sqrt{b_{1}-1}}\right)\Bigg],\label{2111cb}
	\end{align}
	\normalsize
	and this (\ref{2108}) is the massive $F(T)$ form that accepts a wormhole 
solution. If we assume that 	
$r(T)=r_0\,\sqrt{-T}$ then we find \begin{align}
		\mathcal{F}_{-1}(T)=&\frac{2}{3r_0\,(-T)^{3/2}}, \quad  
\mathcal{F}_{-2}(T)=\frac{1}{2r_0^2\,T^2},
	\end{align}
	which leads to  
	\begin{align}
		b(T)=\frac{r_0^3}{2}\,(-T)^{3/2}\left[T+\delta\,\delta_2\,2\sqrt{2}r_0^{-1}\right].
	\end{align}
	 and
	\begin{align}\label{2108d}
		F(T) = \kappa \left(1+3\alpha_{CF}\right)\rho_{0}+F_1\,T+\kappa\,m^2\,\Bigg[\frac{2\tilde{c}_1}{r_0\,\sqrt{-T}}+\frac{\tilde{c}_2}{r_0^2\,(-T)}\Bigg] .
	\end{align}

\subsection{Massive gravity with uniform pressure case}

Setting $\Phi(r)=\Phi_0=const.$, eqns \eqref{1502}-\eqref{1504} 
can be simplified as  
\begin{align}
&	2\kappa\left[\alpha_{CF}\,\rho_{CF}+m^2P_0\right]= 
F-2\left(T+\frac{\delta\,\delta_2\sqrt{-2T}}{r(T)}\right)\,F_T \label{2202}
	\\	
&\partial_r\left(\ln\,F_{T}\right)\,\left(T+\frac{\delta\,\delta_2\sqrt{-2T}}{ 
r(T)}\right) 
=-\partial_r\left(\frac{T}{2}+\delta\,\delta_2\,\sqrt{-2T}r^{-1}(T)\right), 
\label{2204}
\end{align}
where $\alpha_{CF}=const.$ and  $\rho_{CF} \neq 0$.
In this case, assuming   $r(T)=r_0\, \sqrt{-T}$ we find 
	\begin{align}\label{2254}
		F(T)=-2\kappa \tilde{\Lambda}_0 +2F_0 
\,\sqrt{T+\sqrt{2}\delta\delta_2\,r_0^{-1}}  . 
	\end{align}
	with $\tilde{\Lambda}_0=-\alpha_{CF}\,\rho_{CF}-m^2P_0$, and where $P_0$ 
and $\alpha_{CF}$ are negative.

	\section{Logarithmic redshift function  solutions}\label{sect5}
	
In this section we examine the subclass of wormhole solutions that have a
logarithmic redshift function.  Setting   
$\Phi(r)=\Phi_0+a\,\ln (r)$, i.e. $A_1=a_0\,r^a$  
\cite{SSpaper,nonvacSSpaper,scalarfieldSSpaper}, and leaving   $A_2$ in the 
same form, eqn  \eqref{1405} becomes 
	\begin{align}
		T(r) = & -\frac{2}{r^2}\left[2(1+a)\left(1+\delta\sqrt{1-\frac{b(r)}{r}}\right)-(1+2a)\frac{b(r)}{r}\right]. \label{3002}
	\end{align}
In order to proceed, we need to set the $b(r)$ or $r(T)$ function. { Equation \eqref{3002} was used as
a pure power-law ansatz in recent works  \cite{SSpaper,nonvacSSpaper,scalarfieldSSpaper}. The only difference is the $A_2(r)$ component where we use in fact a superposition or an infinite sum of power-law terms as $A_2(r)=\left(1-\frac{b(r)}{r}\right)^{-1/2} =\sum_{p\geq0}\,\frac{(2p)!}{4^p(p!)^2}\left(\frac{b(r)}{r}\right)^p$. The fundamental expression of the torsion scalar was found   using   \eqref{torsionten}--\eqref{torsionscal}. The $\delta=\pm 1$ parameter for the spin-connection defined by   \eqref{1103} essentially impacts the inertial effects without a significant impact on the gravitational effects of the physical system. We then substitute   Eqns \eqref{1100}-\eqref{1101} for the coframe/spin-connection pair with the appropriate redshift for a pure power-law $A_1$ as used in   \cite{SSpaper,nonvacSSpaper,scalarfieldSSpaper}.}

\subsection{General massive gravity  case}

As we described in subsection \ref{gencase}, we focus on the case 
where   $c_1\gg c_2$ case and $K_3(r)=\frac{r(T)}{\tilde{C}}$. 

\subsubsection { $a=-2$ case}

We first 
examine the $a=-2$ case,  obtaining
	\begin{align}
		&\kappa \left(\rho_{CF}(0)\,r^{\frac{2(1+\alpha_{CF})}{\alpha_{CF}}}+m^2\tilde{c}_1r(T)\right)
		= -\frac{1}{2}\left[F-TF_T\right]-\frac{2}{r}\partial_r\left(F_{T}\right)\,\left[\delta\left(1-\frac{b(r)}{r}\right)^{1/2}+\left(1-\frac{b(r)}{r}\right)\right]+{F_T}\,\frac{b'(r)}{r^2} ,\label{3302}
		\\
		&{\kappa m^2\tilde{c}_1}r^2= 2\partial_r\left(F_{T}\right)\,\left[\left(1-\frac{b(r)}{r}\right)-\delta\left(1-\frac{b(r)}{r}\right)^{1/2}\right]-F_T\Bigg[\frac{\left(16+b'(r)\right)}{r}-\frac{15b(r)}{r^2}\Bigg]. \label{3302unified}
	\end{align}
	  Assuming $F_T(T(r))=F_T(r(T))=F_T(T)$, we find
	\begin{align}\label{3302solution}
		F_T(T)=& u^{-1}(T)\Bigg[\frac{\kappa m^2\tilde{c}_1}{2}\int_{r(T)}\,dr'\,\frac{u(T(r'))\,r'^2}{\left(1-\frac{b(r)}{r}\right)^{1/2}\left(\left(1-\frac{b(r)}{r}\right)^{1/2}-\delta\right)}+C_1\Bigg] ,
	\nonumber\\
	u(T) =u(r(T))=& 
\exp\left[-\frac{1}{2}\int_{r(T)}\,\frac{dr'}{r'}\,\frac{\left(16+b'(r)-\frac{
15b(r)}{r}\right)}{\left(1-\frac{b(r)}{r}\right)^{1/2}\left(\left(1-\frac{b(r)}{
r}\right)^{1/2}-\delta\right)}\right],
	\end{align}
	and thus eqn \eqref{3302} leads to the     $F(T)$ 
solution  
    \small
	\begin{align}\label{3302solutionfin}
		F(T)=-2\kappa 
\left[\rho_{CF}(0)\,(r(T))^{\frac{2(1+\alpha_{CF})}{\alpha_{CF}}}+m^2\tilde{c}
_1r(T)\right] 
+\left[T+\frac{2b'(r(T))}{r^2(T)}\right]F_T-\frac{4\partial_r\left(F_{T}\right)} 
{r(T)}\left[\delta\left(1-\frac{b(r(T))}{r(T)}\right)^{1/2}+\left(1-\frac{ 
b(r(T))}{r(T)}\right)\right].
	\end{align}
    \normalsize
    
Considering  $r(T)=r_0\,(-T)^{-1/2}$ we acquire
		\begin{align}\label{3306shape}
			{b(T)}=\frac{2r_0}{3\sqrt{-T}}\Bigg[\left[\frac{r_0^2}{4} 
+\frac{2}{3}\right]\pm \Bigg[\frac{4}{9}-\frac{r_0^2}{6}\Bigg]^{1/2}\Bigg]= 
b_0\,(-T)^{-1/2}.
		\end{align}
		Additionally, eqn \eqref{3302unified} gives
			\begin{align}\label{3306unified}
			Q_0\,(-T)^{-5/2}=&  F_{TT}+\frac{P_0}{(-T)}\,F_T,
		\end{align}
		with 
$P_0=\frac{4\left(b_0-r_0\right)}{r_0\,\left(1-\frac{b_0}{r_0}\right)^{1/2}\left
[\left(1-\frac{b_0}{r_0}\right)^{1/2}-\delta\right]}$ and $Q_0=\frac{\kappa 
m^2\tilde{c}_1r_0^3}{4\left(1-\frac{b_0}{r_0}\right)^{1/2}\left[\left(1-\frac{
b_0}{r_0}\right)^{1/2}-\delta\right]}$. 
Thus, for the general case where $P_0 \neq \left\lbrace -\frac{3}{2},\,-1 
\right\rbrace$, we find that $F_T(T)= 
\frac{Q_0}{P_0+\frac{3}{2}}\,(-T)^{-3/2}+C_1\,T^{P_0}$, where $C_1$ is an 
integration constant, and the final solution is:
		\begin{align}\label{3308}
			F(T)=& -2\kappa \tilde{\rho}_{CF}(0)\,(-T)^{-\frac{(1+\alpha_{CF})}{\alpha_{CF}}}-A\,(-T)^{-1/2}
			+C_1\,T^{P_0+1},
		\end{align}
		where $A=2\kappa m^2\tilde{c}_1r_0+\frac{Q_0\left(1+\frac{2b_0}{r_0^3}\right)}{\left(P_0+\frac{3}{2}\right)}+\left(\delta\left(1-\frac{b_0}{r_0}\right)^{1/2}+\left(1-\frac{b_0}{r_0}\right)\right)\frac{12Q_0}{r_0^2\left(P_0+\frac{3}{2}\right)}$.

	\subsubsection{$a=-1$ case}
	
	For the $a=-1$ cases of subsection \ref{gencase},  eqns \eqref{1402} 
and   \eqref{1404}-\eqref{1403}   become:	
	\begin{align}
		&\kappa \left(\rho_{CF}(0)\,r^{\frac{(1+\alpha_{CF})}{\alpha_{CF}}}+m^2\frac{\tilde{c}_2}{r(T)}\right)
		= -\frac{1}{2}\left[F-TF_T\right]-\frac{2}{r}\partial_r\left(F_{T}\right)\,\left[\delta\left(1-\frac{b(r)}{r}\right)^{1/2}+\left(1-\frac{b(r)}{r}\right)\right]+{F_T}\,\frac{b'(r)}{r^2} ,\label{3502}
		\\
		&-\kappa m^2{\tilde{c}_2}= \frac{F_T}{r(T)}\left[3-\frac{2b(r)}{r}\right] +\partial_r\left(F_{T}\right)\,\left(1-\frac{b(r)}{r}\right)^{1/2}\left[\delta-\left(1-\frac{b(r)}{r}\right)^{1/2}\right] .\label{3502unified}
	\end{align}
  The solution of   \eqref{3502unified} is 
	\begin{eqnarray}\label{3502solution}
	&&	F_T(T)=w^{-1}(T)\left[-\kappa m^2{\tilde{c}_2}\int_{r(T)}\,dr'\, 
\frac{w(r')}{\left(1-\frac{b(r)}{r}\right)^{1/2}\left[\delta-\left(1-\frac{b(r)}
{r}\right)^{1/2}\right]}\,+C_2\right]
	 \\
&&w(T)=w(r(T))=\exp\left[\int_{r(T)}\,\frac{dr'}{r'}\,\frac{\left(3-\frac{2b(r')
} {r'}\right)}{\left(1-\frac{b(r')}{r'}\right)^{1/2}\left[\delta-\left(1-\frac{ 
b(r')}{r'}\right)^{1/2}\right]}\right] ,
\label{wsol222}
	\end{eqnarray}
	and thus from  \eqref{3502}, the general   $F(T)$ solution 
is 
    \small
	\begin{align}\label{3502solutionfinal}
	F(T)=-2\kappa \left[\rho_{CF}(0)\,(r(T))^{\frac{(1+\alpha_{CF})}{\alpha_{CF}}}+m^2\frac{\tilde{c}_2}{r(T)}\right]
		+ \left[T+\frac{2b'(r(T))}{r^2(T)}\right]F_T-\frac{4\partial_r\left(F_{T}\right)}{r(T)}\left[\delta\left(1-\frac{b(r(T))}{r(T)}\right)^{1/2}+\left(1-\frac{b(r(T))}{r(T)}\right)\right].
	\end{align}
    \normalsize
    
    Considering  $r(T)=r_0\,(-T)^{-1/2}$ we acquire  $b(T)= 
\frac{r_0^3}{2}\,(-T)^{-1/2}$, while    (\ref{wsol222}) gives 
$
 w(T)= w_0\,\left(-T\right)^{-w_1} 
$
 with 
$w_1=\frac{\left(3-r_0^2\right)}{2\left(1-\frac{r_0^2}{2}\right)^{1/2}\left[
\delta-\left(1-\frac{r_0^2}{2}\right)^{1/2}\right]}$. Thus, for  $w_1 = 
-\frac{1}{2}$ we find
\begin{align}
			F(T)=&	-2\kappa 
\left[\rho_{CF}(0)\,(-T)^{-\frac{(1+\alpha_{CF})}{2\alpha_{CF}}}+\frac{m^2\tilde
{c}_2}{r_0}\sqrt{-T}\left(1+ \ln(-T)\right)\right]
			-F_3\,\sqrt{-T},
			  \label{3509}
		\end{align}
while for the general case $w_1 \neq -\frac{1}{2}$
we   find 	 
		\begin{align}
			F(T)=&	-2\kappa \left[\rho_{CF}(0)\,(-T)^{-\frac{(1+\alpha_{CF})}{2\alpha_{CF}}}+\frac{m^2\tilde{c}_2}{r_0}\left(\frac{2w_1-1}{2w_1+1}\right)\sqrt{-T}\right]
			-F_2\left(-T\right)^{w_1+1}.
			. \label{3508}
		\end{align}
Hence, in the limit  $w_1\gg 1$ with $\alpha_{CF}= \Delta 
\alpha_{CF}\,\rightarrow\,0$ (ordinary matter limit) we obtain
				\begin{align}
			F(T)\approx &		-2\kappa \rho_{CF}(0)\,(-T)^{-\frac{1}{2\Delta \alpha_{CF}}}
			-F_2\left(-T\right)^{w_1}
			. \label{3508large}
		\end{align}

\subsection{Massive gravity with uniform pressure case}

\subsubsection { $a=-2$ case}
	
 In this case   the  FEs   \eqref{1503}-\eqref{1504} become
	\small
	\begin{align}
		&\kappa \left(\rho_{CF0}\,r^{-\frac{a\left(1+\alpha_{CF}\right)}{\alpha_{CF}}}+\frac{3m^2}{2}\,K_3\left(c_1+2c_2K_3\right)\right)
		= -\frac{1}{2}\left[F-TF_T\right]-\frac{2}{r}\partial_r\left(F_{T}\right)\left[\delta\left(1-\frac{b(r)}{r}\right)^{1/2}+\left(1-\frac{b(r)}{r}\right)\right]+{F_T}\,\frac{b'(r)}{r^2} ,\label{3402}
		\\
		&\partial_r\left(\ln\,F_{T}\right)\,\left(1-\frac{b(r)}{r}\right)^{1/2}\left[\delta+\left(a+1\right)\left(1-\frac{b(r)}{r}\right)^{1/2}\right]=\Bigg[\frac{2a-a^2}{r}+\frac{b(r)}{r^2}\left(a^2-3a-1\right)-\frac{b(r)(1-a)-rb'(r)(a+1)}{2r^2}\Bigg] . \label{3404}
	\end{align}
	\normalsize
	The solution of eqn \eqref{3404} is  
	\begin{align} F_T(T)=F_T(0)\,\exp\Bigg[\int_{r(T)}\,dr\,\frac{\Bigg[\frac{2a-a^2}{r}+\frac{b(r)}{r^2}\left(a^2-3a-1\right)-\frac{b(r)(1-a)-rb'(r)(a+1)}{2r^2}\Bigg]}{\left(1-\frac{b(r)}{r}\right)^{1/2}\left[\delta+\left(a+1\right)\left(1-\frac{b(r)}{r}\right)^{1/2}\right]}\Bigg]. \label{3405}
	\end{align}
	Hence, assuming  $r(T)=r_0\,(-T)^{-1/2}$, then  for $c_1\ll c_2$
	we find  
	\begin{align}
				F(T) =& -2\kappa 
\left(\rho_{CF0}\,(-T)^{\frac{a\left(1+\alpha_{CF}\right)}{2\alpha_{CF}}}
+3m^2\tilde{c}_2(-T)^{2a}\right)-F_0\left(1-\frac{2b_0}{r_0^3}\right)(-T)^{y+1}
+R_0\,(-T)^{y+1}, \label{3409}
			\end{align} 
	while for  $c_1\gg c_2$   we find:
	 		\begin{align}
				F(T) =&	-2\kappa 
\left(\rho_{CF0}\,(-T)^{\frac{a\left(1+\alpha_{CF}\right)}{2\alpha_{CF}}}+\frac{
3m^2\tilde{c}_1}{2}(-T)^{\frac{5a}{4}}\right)-F_0\left(1-\frac{2b_0}{r_0^3}
\right)(-T)^{y+1}+R_0\,(-T)^{y+1}, \label{3410}
			\end{align} 
  		where 
$R_0=\frac{8y}{r_0^2}F_0\left[\delta\left(1-\frac{b_0}{r_0}\right)^{1/2}
+\left(1-\frac{b_0}{r_0}\right)\right]$.

	\subsubsection{$a=-1$ case}
	
	In this case the FEs \eqref{1503}-\eqref{1504} become
	\begin{align}
		&\kappa \left(\rho_{CF0}\,r^{\frac{\left(1+\alpha_{CF}\right)}{\alpha_{CF}}}+\frac{3m^2}{2}\,K_3\left(c_1+2c_2K_3\right)\right)
		= -\frac{1}{2}\left[F+2TF_T\right]-\frac{2}{r}\partial_r\left(F_{T}\right)\left[\delta\left(1+\frac{r^2(T)\,T}{2}\right)^{1/2}+\left(1+\frac{r^2(T)\,T}{2}\right)\right] ,\label{3702}
		\\
		& \partial_r\left(\ln\,F_{T}\right)= -\delta\,r(T) 
\left(T+\frac{3}{r^2(T)}\right)\left(1+\frac{r^2(T)\,T}{2}\right)^{-1/2}.
\label{3704} 
	\end{align}
		The solution of eqn \eqref{3704} is 
	\begin{align}
		F_T(T)=	\exp\left[-\delta\,\int_{r(T)}\,dr'\,r'\left(T(r')+\frac{3}{r'^2}\right)\left(1+\frac{r'^2\,T(r')}{2}\right)^{-1/2}\right] \label{3705}	.
	\end{align}
	Hence, assuming  $r(T)=r_0\,(-T)^{-1/2}$, then  for $c_1\ll c_2$ we find 
	\begin{align}
				F(T)=&	-2\kappa 
\left(\rho_{CF0}\,(-T)^{-\frac{\left(1+\alpha_{CF}\right)}{2\alpha_{CF}}}
+3m^2\tilde{c}_2(-T)^{-2}\right)+2F_0\,(-T)^{y_2+1} -R_1\,(-T)^{y_2+1}, 
\label{3709}
			\end{align}
		while for  $c_1\gg c_2$   we have:
		\begin{align}
				F(T)=&	-2\kappa 
\left(\rho_{CF0}\,(-T)^{-\frac{\left(1+\alpha_{CF}\right)}{2\alpha_{CF}}}+\frac{
3m^2\tilde{c}_1}{2}\,(-T)^{-5/4}\right)+2F_0(-T)^{y_2+1}-R_1\,(-T)^{y_2+1} , 
\label{3710}
			\end{align}
		where 
$R_1=\frac{8y_2}{r_0^2}F_0\left[\delta\left(1-\frac{r_0^2}{2}\right)^{1/2}
+\left(1-\frac{r_0^2}{2}\right)\right]$.

\section{Power-law redshift   function solutions}\label{sect6}

In this section we examine the subclass of wormhole solutions that have a
power-law redshift function. Setting   $\Phi(r)=\Phi_0+a_1\,r^a$, i.e. 
 $A_1=a_0\,\exp(a_1\,r^a)$, and  leave $A_2$ in the same form, eqn  
\eqref{1405} becomes
	\begin{align}
		T(r) =  -\frac{2}{r^2}\left(\delta+\left(1-\frac{b(r)}{r}\right)^{1/2}\right)\left(\delta+\left(1+2a\,a_1 r^a\right)\left(1-\frac{b(r)}{r}\right)^{1/2}\right). \label{4001}
	\end{align}
	 In order to proceed we need to set the $b(r)$ 
or $r(T)$ function .

\subsection{General massive gravity  case}

\subsubsection{$a=1$ and $a_1<0$ case}

In this case eqns   \eqref{1403}-\eqref{1404} become
\small
\begin{align}
	&\kappa \left(\rho_{CF}+m^2K_3\left(c_1+c_2\,K_3\right)\right)
	= -\frac{1}{2}\left[F-TF_T\right]-T^2F_{TT}\,{\left(1-\frac{b_0}{2}\right)}\left[\delta+\left(1-\frac{b_0}{2}\right)\right]-{TF_T}\,\frac{b_0}{4} ,\label{4102}
	\\
	& \kappa m^2K_3\left(\frac{c_1}{2}+c_2\,K_3\right) = -T^2F_{TT}\,\frac{\left(1-\frac{b_0}{2}\right)}{2}\left[\delta+\left(1-2|a_1| (-T)^{-1/2} \right)\left(1-\frac{b_0}{2}\right)\right] +F_T\Bigg[\frac{|a_1|}{2}(b_0-1)\sqrt{-T}  +\frac{b_0}{4}\,T+|a_1|^2 (b_0-1)\Bigg] . \label{4103}
\end{align}
\normalsize
 Hence, for  $c_1\ll c_2$ we extract the solution
	\small
	\begin{align}\label{4105}
		F(T)=& -2\kappa \left(\rho_{CF}(0)\,\exp\left(\frac{2|a_1|\left(1+\alpha_{CF}\right)}{\alpha_{CF}\sqrt{-T}}\right)+m^2\tilde{c}_2\, (-T)\exp\left(\frac{2|a_1|}{\sqrt{-T}}\right) \right)+\frac{F_T(0)}{V(T)}\Bigg[C_1(-T)+\frac{8\kappa m^2\tilde{c}_2}{A_2}\,{(-T)^{1-A_2/2}} 
		\nonumber\\
		&\,+\Bigg[\frac{V_T(T)}{V(T)}\Bigg[C_1(-T)^2-\frac{16\kappa 
m^2\tilde{c}_2}{A_2}\,{(-T)^{2-A_2/2}}\Bigg]+{8\kappa 
m^2\tilde{c}_2}\,{(-T)^{1-A_2/2}}\Bigg]\left[\delta+\left(1-\frac{b_0}{2}
\right)\right]\Bigg],
	\end{align}
	\normalsize
while  for  $c_1\gg c_2$ we extract the solution
	\small
	\begin{align}\label{4106}
		F(T)=& -2\kappa \left(\rho_{CF}(0)\,\exp\left(\frac{2|a_1|\left(1+\alpha_{CF}\right)}{\alpha_{CF}\sqrt{-T}}\right)+m^2\tilde{c}_1\,\sqrt{-T}\exp\left(\frac{2|a_1|}{\sqrt{-T}}\right)\right)+\frac{F_T(0)}{V^(T)}\Bigg[C_1(-T)+\frac{4\kappa m^2\tilde{c}_1}{\left(A_2+1\right)}\,{(-T)^{\left(1-A_2\right)/2}}
		\nonumber\\
		&\,+\Bigg[\frac{V_T(T)}{V(T)}\Bigg[C_1(-T)^2-\frac{8\kappa m^2\tilde{c}_1}{\left(A_2+1\right)}\,{(-T)^{\left(3-A_2\right)/2}}\Bigg]
		+{4\kappa m^2\tilde{c}_2}\,{(-T)^{(1-A_2)/2}}\Bigg]\left[\delta+\left(1-\frac{b_0}{2}\right)\right]\Bigg] ,
	\end{align}
	\normalsize
where
\begin{align}
	V(T)=\left[2|a_1|(b_0-2)(-T)^{-1/2}+2(\delta+1)-b_0\right]^{A_1}\,(-T)^{-A_2/2}\,\exp\left(\frac{A_3}{\sqrt{-T}}\right),
\end{align}
$A_1=\frac{\left[4b_0^3-24(\delta+1)b_0^2+8(8+9\delta)b_0-48(\delta+1)\right]}{
(b_0-2)\left(b_0-2(\delta+1)\right)}=1$, 
$A_2=\frac{4b_0}{(b_0-2)\left(b_0-2(\delta+1)\right)}$ and 
$A_3=\frac{8|a_1|(b_0-1)}{(b_0-2)^2}=-2\,|a_1|$.

\subsubsection{$a=2$ and $a_1<0$ case}

In this case,  for  $c_1\ll c_2$ we extract the solution 
	\small
	\begin{align}
		F(T)=&	-2\kappa 	\left(\rho_{CF}(0)\,\exp\left(-\frac{4|a_1|\left(1+\alpha_{CF}\right)}{\alpha_{CF}\,T}\right)+m^2\tilde{c}_2\,(-T)\exp\left(-\frac{4|a_1|}{T}\right)\right)+\frac{F_T(0)}{W(T)}
		\Bigg[C_1\,T+{2\kappa m^2\tilde{c}_2}\,T\,Ei\left(-\frac{B_2}{T^2}\right)
		\nonumber\\
		&\, +\Bigg[\frac{W_T(T)}{W(T)}\Bigg[C_1\,T^2+{4\kappa m^2\tilde{c}_2}\,T^2\,Ei\left(-\frac{B_2}{T^2}\right)\Bigg]
		-{8\kappa m^2\tilde{c}_2}\,T\,\exp\left(\frac{B_2}{T^2}\right)\Bigg]\left[\delta+\left(1-\frac{b_0}{2}\right)\right]\Bigg],
	\end{align}
	\normalsize
	while  for  $c_1\gg c_2$ we extract the solution
 	\small
	\begin{align}
		F(T)=&	-2\kappa 	\left(\rho_{CF}(0)\,\exp\left(-\frac{4|a_1|\left(1+\alpha_{CF}\right)}{\alpha_{CF}\,T}\right)+m^2\tilde{c}_1\,\sqrt{-T}\exp\left(-\frac{4|a_1|}{T}\right) \right)+\frac{F_T(0)}{W(T)}\Bigg[C_1\,T+\frac{\kappa m^2\tilde{c}_1\,T}{\Gamma\left(\frac{3}{4}\right) \left(-B_2\right)^{1/4}}
		\nonumber\\
		&\,\times\,\left[\Gamma\left(\frac{3}{4}\right) \Gamma\left(\frac{1}{4},-\frac{B_2}{T^2}\right)-\sqrt{2}\pi\right] +\Bigg[\frac{W_T(T)}{W(T)}\Bigg[C_1\,T^2-\frac{2\kappa m^2\tilde{c}_1\,T^2}{\Gamma\left(\frac{3}{4}\right) \left(-B_2\right)^{1/4}}\left[\Gamma\left(\frac{3}{4}\right) \Gamma\left(\frac{1}{4},-\frac{B_2}{T^2}\right)-\sqrt{2}\pi\right]\Bigg]
		\nonumber\\
		&\,+{4\kappa m^2\tilde{c}_1}\, T^{1/2} \exp\left(\frac{B_2}{T^2}\right)\Bigg]\left[\delta+\left(1-\frac{b_0}{2}\right)\right]\Bigg] ,
	\end{align}
	\normalsize
where 
\begin{align}
	W(T) = \left[\left(2(\delta+1)-b_0\right)+{8|a_1|}\left(b_0-2\right)\,(-T)^{-1}\right]^{B_1} \exp\Bigg[(-T)^{-1}\left(B_2(-T)^{-1}+B_3\right)\Bigg] ,
\end{align}
 $B_1=\frac{\left[-b_0^3+2b_0^2(1-2\delta)+4b_0(3+4\delta)-16(\delta+1)\right]}{8|a_1|(b_0-2)^4}=1$, $B_2=\frac{8|a_1|\left(b_0^2-3b_0+2\right)}{(b_0-2)^3}$ and $B_3=\frac{2\left(2(\delta+1)-b_0(1+2\delta)\right)}{(b_0-2)^3}=-4|a_1|$. For $B_1 \neq 1$ and/or $B_3 \neq -4|a_1|$, the teleparallel solutions will be described by more complex expressions.

\subsection{Massive gravity with uniform pressure case}
	
	\subsubsection{$a=1$ and $a_1<0$ case}
	
	In this case,  for  $c_1\ll c_2$ we find the solution  
		\begin{align}
			F(T)=&	-2\kappa \left(\rho_{CF}(0)\,\exp\left(\frac{2|a_1|\left(1+\alpha_{CF}\right)}{\alpha_{CF}\sqrt{-T}}\right)+{3m^2\tilde{c}_2}\,\exp\left(\frac{8|a_1|}{\sqrt{-T}}\right)\right)+F_{3G}(T) ,
		\end{align}
			while  for  $c_1\gg c_2$ we find the solution		 
		\begin{align}
	F(T)=&	-2\kappa 
\left(\rho_{CF}(0)\,\exp\left(\frac{2|a_1|\left(1+\alpha_{CF}\right)}{\alpha_{CF
}\sqrt{-T}}\right)+\frac{3m^2\tilde{c}_1}{2}\,\exp\left(\frac{5|a_1|}{\sqrt{-T}}
\right)\right)+F_{3G}(T) ,
		\end{align}
				where
		\small
			\begin{align}
			F_{3G}(T)=&	-F_T(0)\, (-T)^{P_3+1} \exp\Bigg[\frac{R_3}{\sqrt{-T}}\Bigg]\left(\left(\delta +1-\frac{b_0}{2}\right) \sqrt{-T}-2|a_1|\left(1-\frac{b_0}{2}\right)\right)^{Q_3}\left(1-\frac{b_0}{2}\right)\Bigg[1+
			\Bigg[2P_3+{R_3}(-T)^{-1/2}
			\nonumber\\
			&\quad -\frac{Q_3 \left(\delta +1-\frac{b_0}{2}\right)\,\sqrt{-T}}{\left(\left(\delta +1-\frac{b_0}{2}\right) \sqrt{-T}-2|a_1|\left(1-\frac{b_0}{2}\right)\right)}\Bigg]
			\left[\delta+\left(1-\frac{b_0}{2}\right)\right] \Bigg]
			\nonumber\\	
			P_3=&-\frac{\left(- \left(\frac{1}{2}+|a_1|\left(b_0-1\right)\right) \left(\delta +1-\frac{b_0}{2}\right)+8|a_1|\left(1-\frac{b_0}{2}\right)\right) \ln\! \left(-T\right)}{2|a_1|\left(1-\frac{b_0}{2}\right)^{3}}
			\nonumber\\ 
			Q_3=&-\frac{\Bigg[\left(1+2|a_1| \left(b_0-1\right)\right) \,\left(\delta +1-\frac{b_0}{2}\right)^{2}-16 |a_1|\left(1-\frac{b_0}{2}\right) \left(\delta +1-\frac{b_0}{2}\right) -b_{0} |a_1| \left(1-\frac{b_0}{2}\right)^{2}\Bigg]}{2|a_1|\left(1-\frac{b_0}{2}\right)^{3} \left(\delta +1-\frac{b_0}{2}\right)},
			\nonumber\\
			R_3=& -\frac{\left(1+2|a_1| \left(b_0-1\right)\right)}{\left(1-\frac{b_0}{2}\right)^2}					 .
		\end{align}	
		\normalsize

	\subsubsection{$a=2$ and $a_1<0$ case}
	
		In this case,  for  $c_1\ll c_2$ we extract the solution   
		\begin{align}
			F(T)=&	-2\kappa \left(\rho_{CF}(0)\,\exp\left(-\frac{4|a_1|\left(1+\alpha_{CF}\right)}{\alpha_{CF}\,T}\right)+{3m^2\tilde{c}_2}\,\exp\left(-\frac{16|a_1|}{T}\right) \right) +F_{4G}(T) , 
		\end{align}
while  for  $c_1\gg c_2$ we extract the solution		 
		\begin{align}
		F(T)=&	-2\kappa 	\left(\rho_{CF}(0)\,\exp\left(-\frac{4|a_1|\left(1+\alpha_{CF}\right)}{\alpha_{CF}\,T}\right)+\frac{3m^2\tilde{c}_1}{2}\,\exp\left(-\frac{10|a_1|}{T}\right)\right) +F_{4G}(T) ,  
		\end{align}
	where
	\small
\begin{align}
	F_{4G}(T)=& F_T(0)\,T^{P_4+1}\,\exp\left(\frac{R_4}{T}\right)\left(\left(\delta+1-\frac{b_0}{2} \right) T+8|a_1|\left(1-\frac{b_0}{2}\right)\right)^{Q_4}\left(1-\frac{b_0}{2}\right)\Bigg[1
	-2\left[\delta+\left(1-\frac{b_0}{2}\right)\right]\Bigg[P_4-\frac{R_4}{T}
	\nonumber\\
	&\quad+\frac{Q_4\left(\delta+1-\frac{b_0}{2} \right) T}{\left(\left(\delta+1-\frac{b_0}{2} \right) T+8|a_1|\left(1-\frac{b_0}{2}\right)\right)}\Bigg]\Bigg] ,
	\nonumber\\
	P_4=& \frac{\left(b_0-1\right)\left(\delta+1-\frac{b_0}{2} \right)}{2\left(1-\frac{b_0}{2}\right)^2} ,
	\quad 
	Q_4= \frac{b_0\left(1-\frac{b_0}{2}\right)^2-\left(b_0-1\right) \left(\delta+1-\frac{b_0}{2} \right)^2}{2\left(\delta+1-\frac{b_0}{2} \right)\left(1-\frac{b_0}{2}\right)^3}, \quad R_4=\frac{4\left(b_0-1\right)|a_1|}{\left(1-\frac{b_0}{2}\right)^2} .
\end{align}
\normalsize

\begin{figure}[!htp]
	\centering
	\includegraphics[width=13.0cm,height=11.0cm]{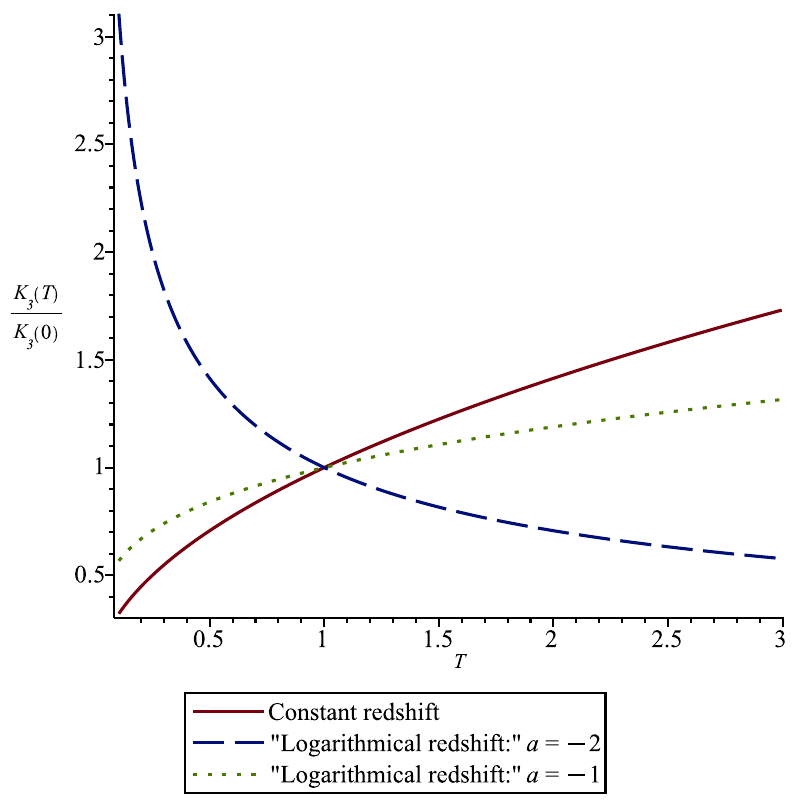} \\
		\includegraphics[width=13.0cm,height=11.0cm]{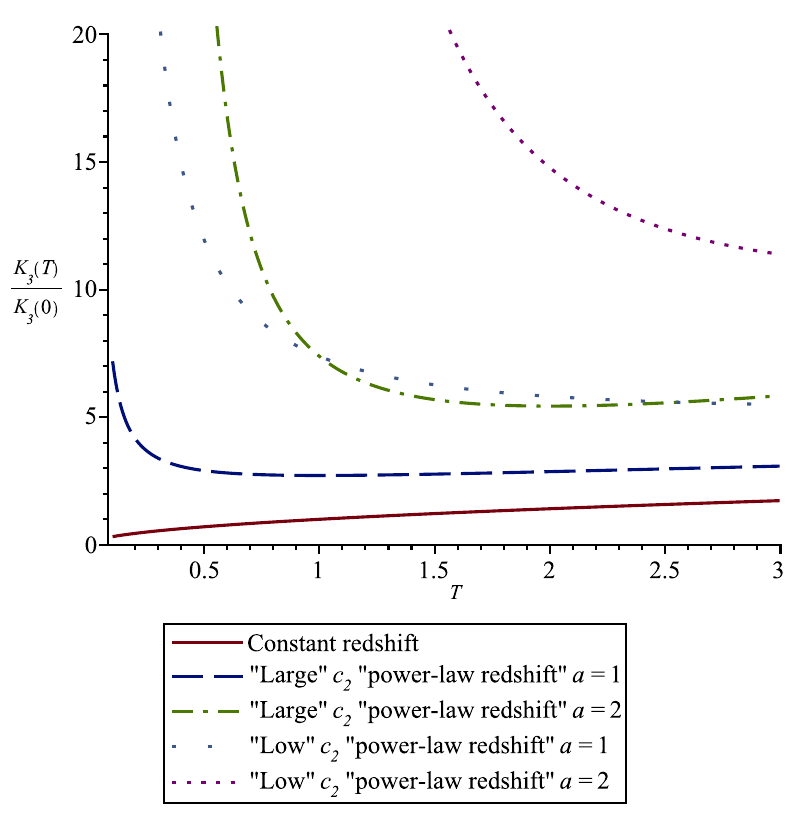}   
		\caption{The $K_3(T)$ massive function versus $T$ of 
constant, $a=-1$ and $-2$ logarithmic and $a=1$ and $2$ power-law redshift 
function cases for general massive sources.}
		\label{figure1}
\end{figure}

\begin{figure}[!htp]
	\centering
	     \includegraphics[width=13.0cm,height=11.0cm]{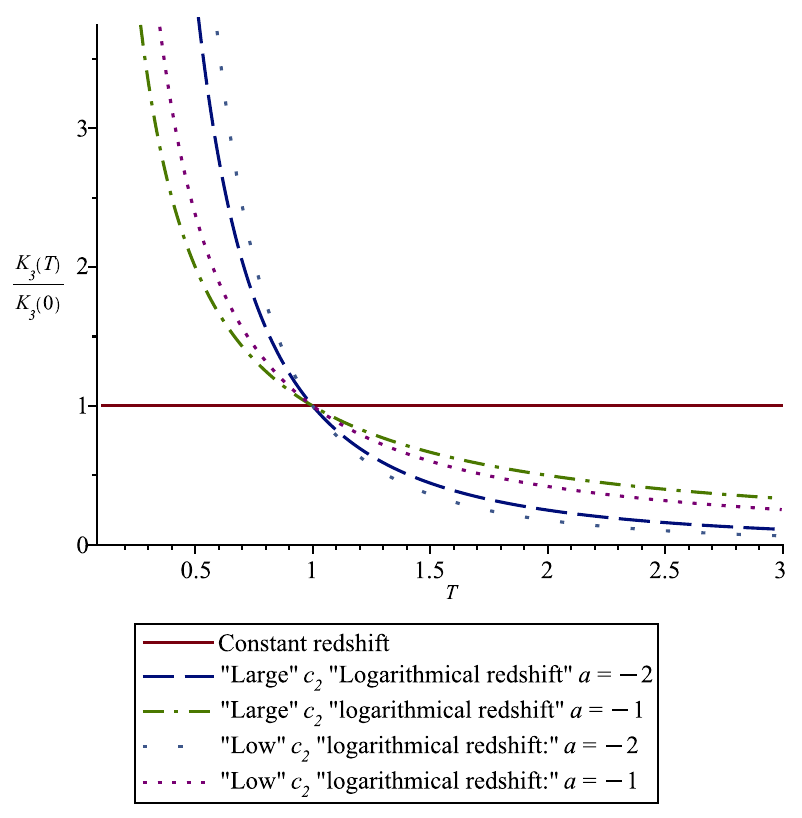}  \\
	       \includegraphics[width=13.0cm,height=11.0cm]{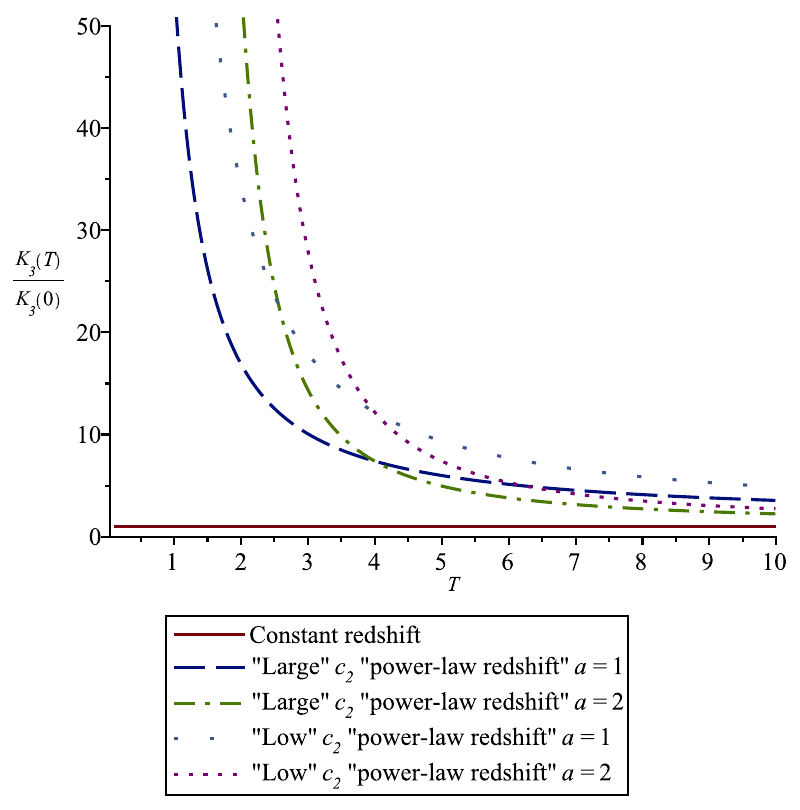} 
		\caption{$K_3(T)$ massive function versus $T$ of 
constant, $a=-1$ and $-2$ logarithmic and $a=1$ and $2$ power-law redshift 
function cases fo uniform pressure  massive sources.}
		\label{figure1b}
\end{figure}

	\section{Physical features and comparisons between cases}\label{sectphys}

In sections \ref{sect4},  \ref{sect5} and \ref{sect6}, we  extracted exact 
wormhole solutions for massive $F(T)$ gravity, imposing the usual cases for 
  the redshift function, such as constant, logarithmical, and power-law. 
There are some common points to be discussed beyond the main differences in 
the new   $F(T)$ solutions and quantities as the massive $K_3(T)$ 
function. In this section we focus  on these common points  by using the 
$\frac{r(T)}{r_0}=\frac{b(T)}{b_0}=(-T)^{-1/2}$ case for simplicity. We could do the same analysis using other cases examined above; however, the 
results and graphs would be qualitatively similar. 

There are some major differences in the physical impacts between 
$\frac{r(T)}{r_0} > 1$ and $\frac{r(T)}{r_0}<1$ cases, 
especially on the wormhole horizons and  on the limits on the values of 
the torsion scalar $T$. The $\frac{r(T)}{r_0}=1$ (i.e. $T=-1$) boundary case is a 
type of physical limit on the size of wormhole horizons. Nevertheless, this last consideration will be useful to extract wormhole solutions for   
more complicated $r(T)$ forms and then for the corresponding $F(T)$ 
solutions { shown in Fig. \ref{figure3}.}

\begin{figure}[!htp]
	\centering
	\includegraphics[width=13.0cm,height=11.cm]{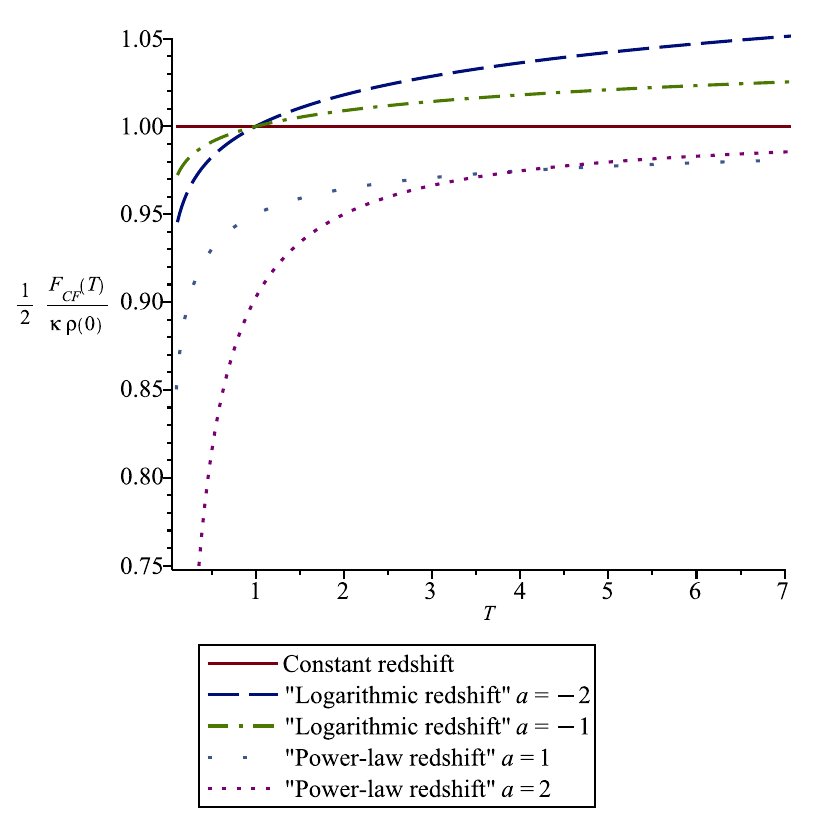} \\
\includegraphics[width=13.0cm,height=11.cm]{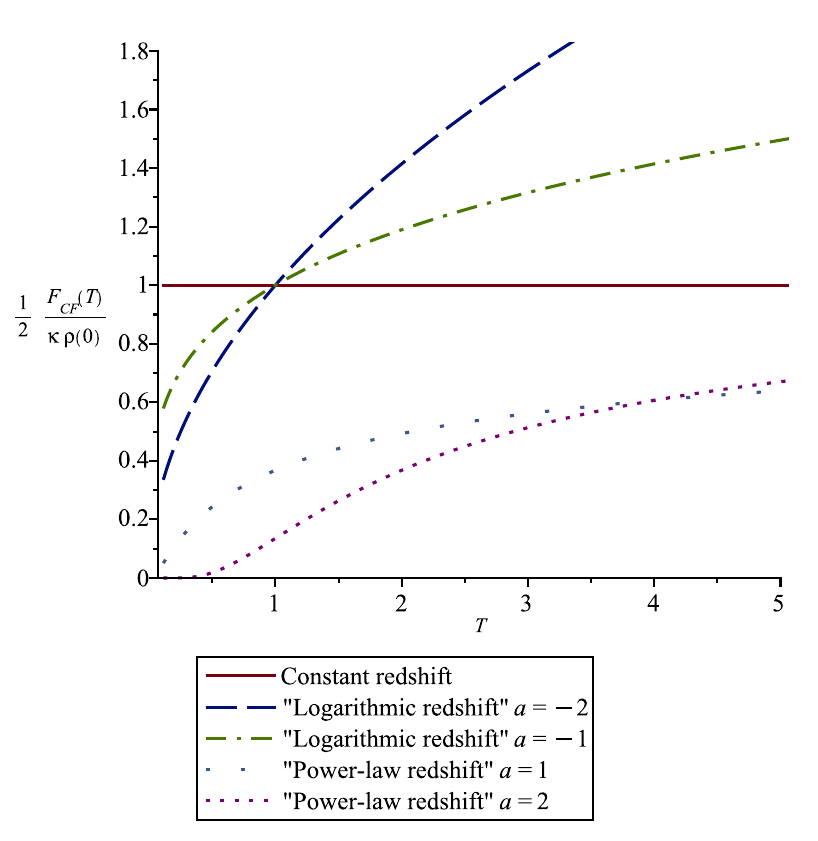} 
\caption{The normalized $F(T)$ versus $T$, of constant, $a=-1$ and $-2$ 
logarithmic and $a=1$ and $2$ power-law redshift functions. Top: 
$\alpha_{CF}\,\rightarrow\,-1$ limit, Bottom: $\alpha_{CF}=-\frac{2}{3}$.}
\label{figure2}
\end{figure}

\begin{figure}[!htp]
	      \includegraphics[width=13.0cm,height=11.cm]{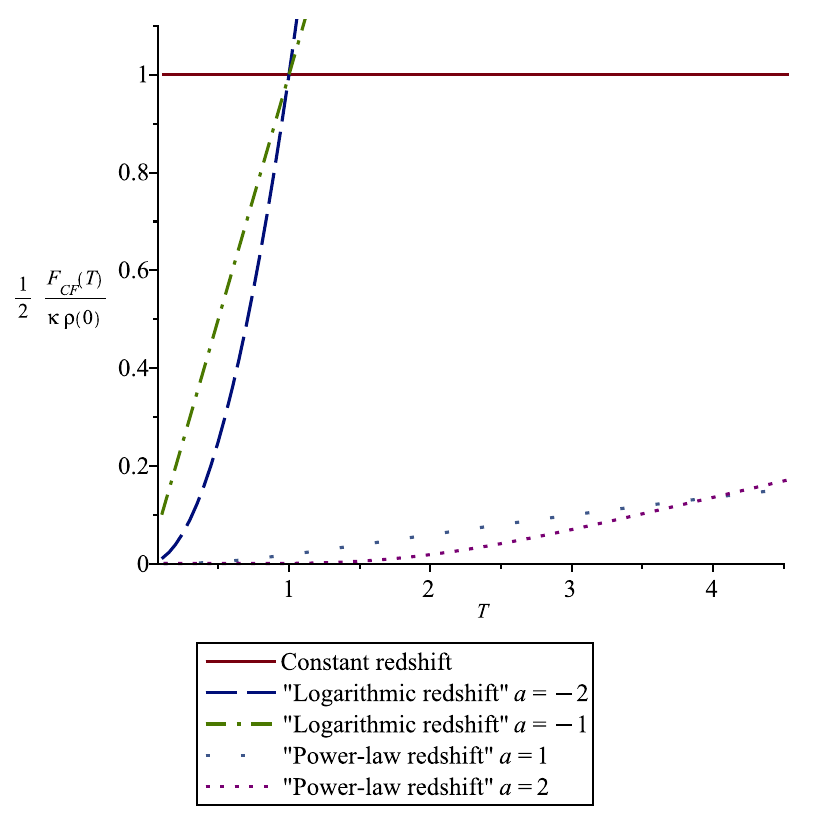}\\
	      \includegraphics[width=13.0cm,height=11.cm]{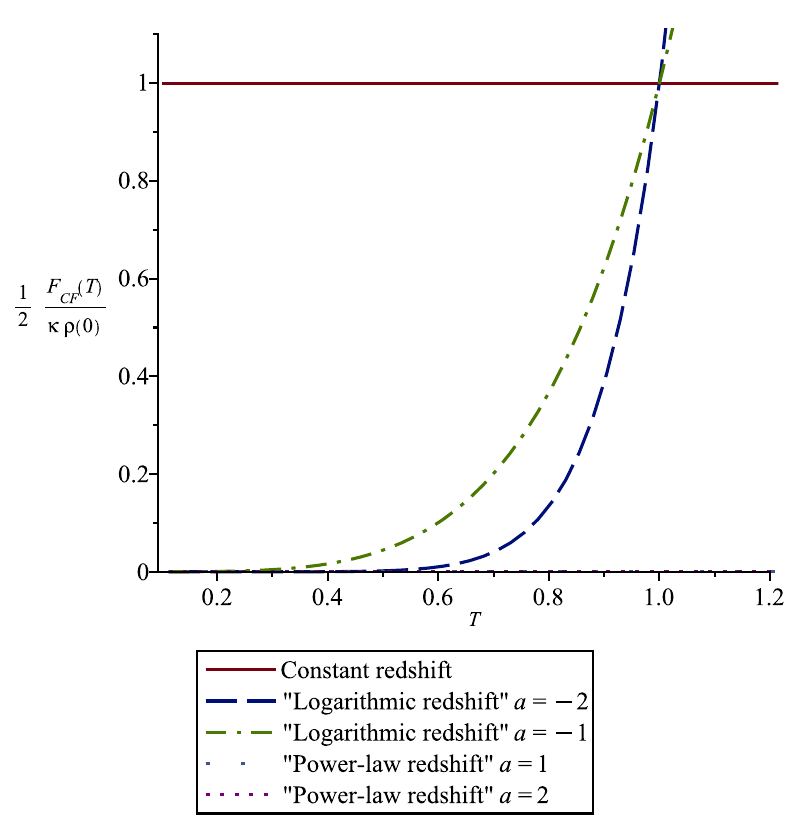}  
	\caption{The normalized $F(T)$ versus $T$, of constant, $a=-1$ and $-2$ 
logarithmic and $a=1$ and $2$ power-law redshift functions. Top: 
$\alpha_{CF}=-\frac{1}{3}$, Bottom: $\alpha_{CF}\,\rightarrow\,0$ limit.}
	\label{figure2a}
\end{figure}

	\begin{figure}[!htp]
	       		\centering
	   \includegraphics[width=8.cm,height=7cm]{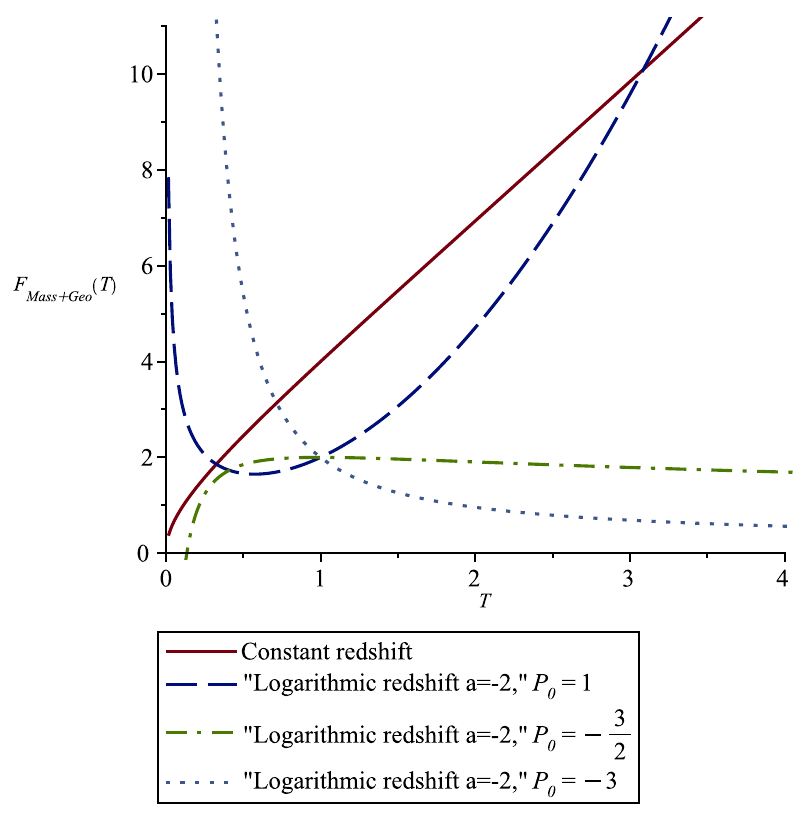}
            \includegraphics[width=8.cm,height=7cm]{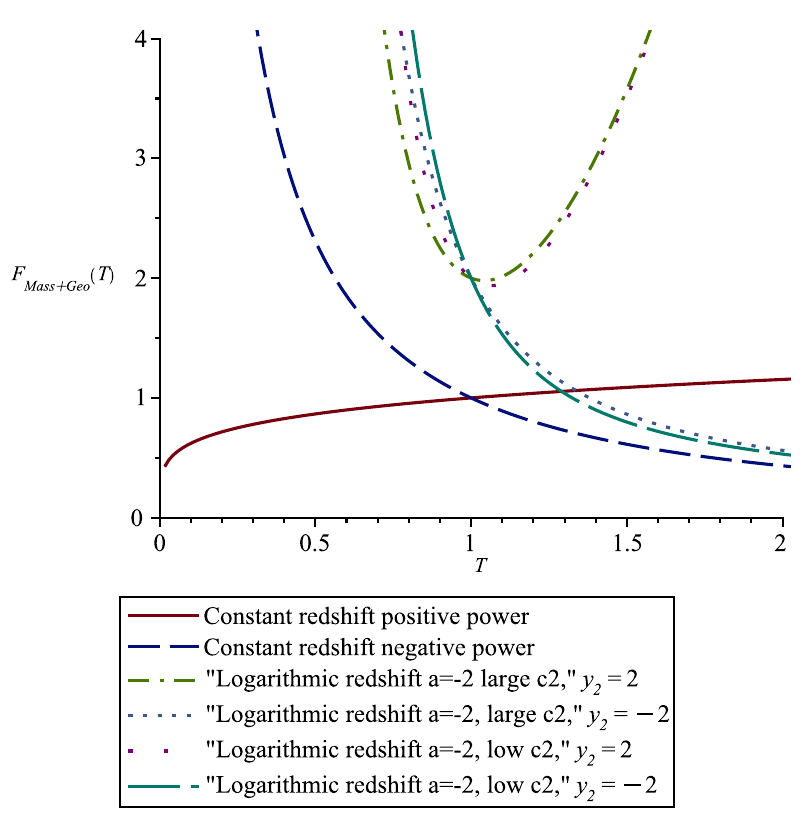}
        \\
	   \includegraphics[width=8.cm,height=7cm]{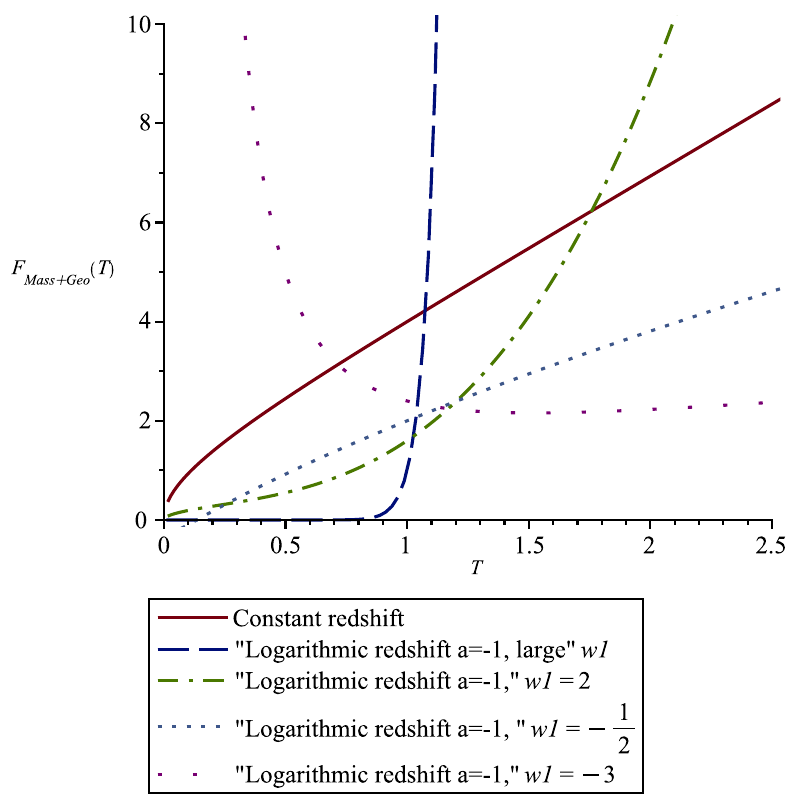} 
           \includegraphics[width=8.cm,height=7cm]{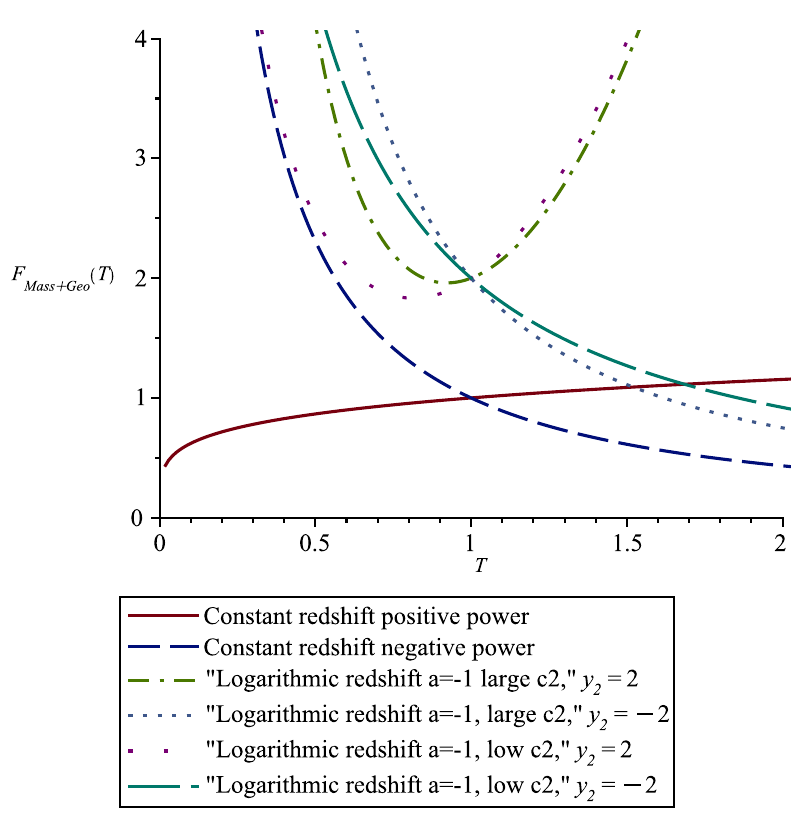}
        \\
        \includegraphics[width=8.cm,height=7cm]{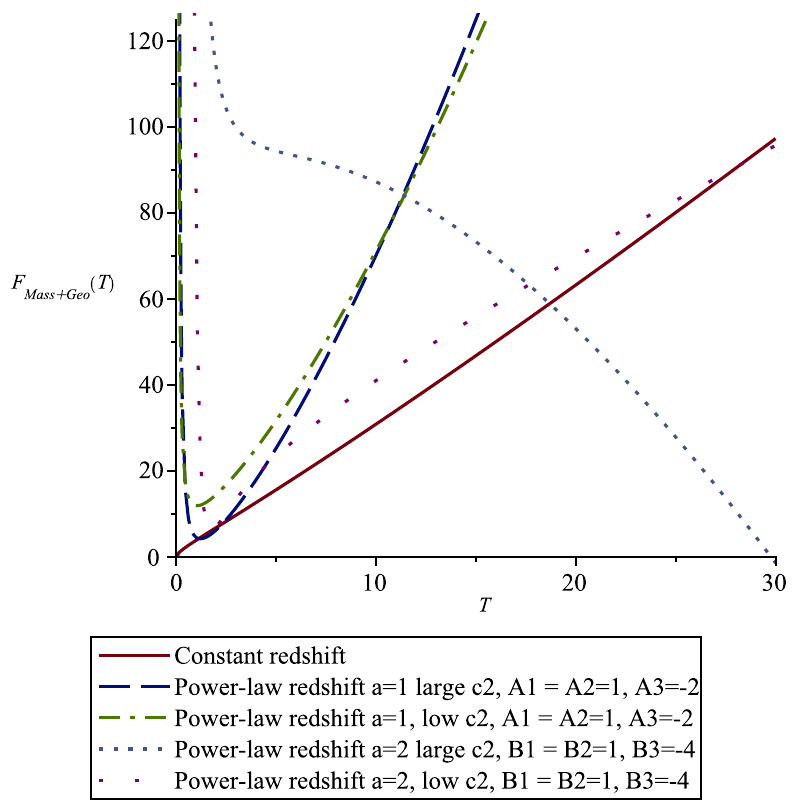} 
           \includegraphics[width=8.cm,height=7cm]{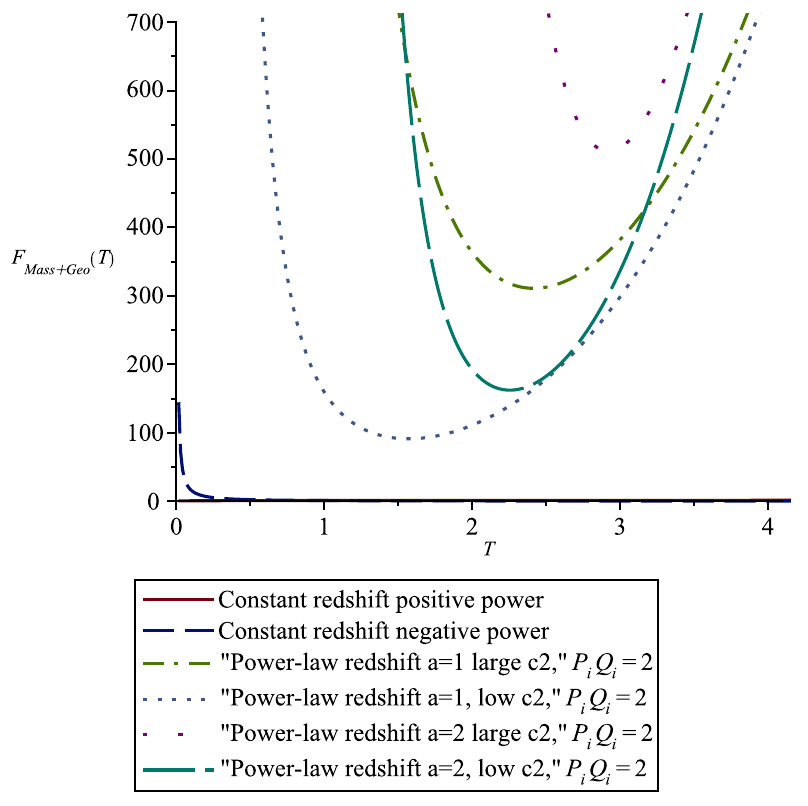}    
		\caption{The massive and geometry contributions to teleparallel 
$F(T)$ solutions versus $T$, of constant, $a=-1$ and $-2$ logarithmic and $a=1$ 
and $2$ power-law redshift functions, for general (left) and uniform 
(right) pressure massive sources.}
		\label{figure3}
	\end{figure}

In Figs. \ref{figure1}--\ref{figure1b} we depict the $K_3$ massive function, expressed as a 
function of $T$. Then in Fig. 
\ref{figure2} we separate the cosmological fluid contribution   
from the massive and geometry terms, and we  highlight the massive term 
contributions from the other terms of the teleparallel $F(T)$ solutions.
Note that the cosmological fluid contributions to solutions for non-interacting 
fluid with the massive gravitons have been studied in 
\cite{nonvacSSpaper,nonvacKSpaper}.   The constant redshift case can be used 
as a reference case for comparison tasks. As we can see in Figs. \ref{figure1} to \ref{figure2a}, the  power-law redshift solutions found in section 
\ref{sect6} lead to the same family of curves of solutions. This is also the case 
for the logarithmic solutions of the uniform pressure case. However, for 
general massive solutions, the source term $K_3(T)$ will be different between the cases $a=-1$ and $a=-2$.

In Fig. \ref{figure3}  we observe that there exist differences between the 
various cases, similarly to Figs. \ref{figure1}--\ref{figure1b}, for the top 
subfigures (general massive source). The comparisons reveal
the direct link 
between the massive source function $K_3(T)$ and the massive term of the $F(T)$ 
solutions. The massive    solution will then affect the corresponding 
$F(T)$ solution for the same geometric term contribution. Under the same 
principle as in Figs. \ref{figure1}--\ref{figure2a}, we use the section 
\ref{sect4}   $F(T)$ solutions as the reference for comparison of 
sections \ref{sect5} and \ref{sect6} new solutions. The general massive 
  $F(T)$ solutions of section \ref{sect5} for $a=-2$ and section 
\ref{sect6} for large $c_2$ and $a=2$ are a bit different from   the rest   
cases, since the $K_3(T)$ function is different from the other cases. In 
particular, for most of the other cases,  we will find that the same families 
of functions are all similar in Fig. \ref{figure3}. The constant 
redshift case is still the reference for comparison, as for Figs. \ref{figure1}--\ref{figure2a}.

In short, the massive $F(T)$ terms 
reflect the  solutions for  the same teleparallel geometry contribution 
including a wormhole spacetime. Each class of massive teleparallel $F(T)$ 
solutions will lead to a new type of wormhole with its specific features. { Each plot in Fig. \ref{figure3} shows the massive term domination shown on Figs. \ref{figure1}-\ref{figure1b},
added by the geometry contributions defined by Eqns. \eqref{1102}-\eqref{1103} pair for appropriate ansatz to new teleparallel $F(T)$ solutions. The massive contribution on each plot essentially reflects the $K_3(T)$ source terms, as expected.} Above all, 
in sections \ref{sect4} to \ref{sect6} we have shown  that a dRGT massive 
source in a cosmological fluid yields several classes of new   
$F(T)$ solutions applicable to any wormhole structure.

\section{Conclusions}\label{sect7}

We constructed and analysed static, spherically symmetric traversable wormhole solutions in a teleparallel gravity setup where the usual $F(T)$ action is augmented by a perturbative dRGT graviton-mass term. Bringing together a torsion-based modification and a nonzero graviton mass couples two independently motivated extensions of general relativity, producing strong-field effects and solution branches absent in either framework alone. Since throat structure, horizon formation, and the effective gravitational stress-energy probe the nonlinear field equations, compact spacetimes such as black holes and wormholes offer natural and stringent tests for the combined dynamics.

We have constructed exact teleparallel solutions by employing the Morris-Thorne wormhole metric and several choices for the redshift function, namely constant, 
logarithmic, and power-law forms. For each case, we derived the corresponding 
FEs and conservation laws, reconstructed the shape function $b(T)$ and 
the allowed forms of teleparallel $F(T)$ solutions { as shown in sections \ref{sect4}-\ref{sect6}. We analyzed the physical behavior near the throat and found that massive source contributions to $F(T)$ reflect the massive conservation laws functions $K_3(T)$. }

The obtained teleparallel solutions satisfy the flaring-out condition and remain free from event horizons. Moreover, in both the general and uniform-pressure massive 
sectors, the effective energy density and pressures can satisfy, or at most 
slightly violate, the ECs. { However, Eqn. \eqref{1407b} might be violated in some cases by its particular form as discussed in section \ref{gencase}. The ECs satisfaction shows that traversable wormholes is clearly able to exist in this framework without the need for exotic matter and dark energy}.

The analysis also revealed that the massive dRGT contribution acts as an 
effective anisotropic source that helps sustain the throat geometry { as we can see in the $K_3(T)$ functions used for each redshift cases.}. Depending 
on the redshift profile { (or $A_1(r)$ frame function)}, the coupling between torsion and the massive term 
modifies the asymptotic behavior of the solutions and determines the range of 
allowed parameters for which the geometry remains regular and asymptotically 
flat. In the limit of vanishing graviton mass, the models reduce smoothly to the 
standard $F(T)$ wormhole configurations { defined in a dark energy universe, confirming the internal consistency of 
the construction and also the recent teleparallel solution classes suitable a dark energy-dominated universe \cite{nonvacSSpaper,scalarfieldSSpaper}.}

Several concrete directions naturally extend this work. First, relaxing spherical symmetry to construct axially symmetric, stationary and rotating solutions (via adapted tetrad ansätze and numerical integration) could reveal families of rotating teleparallel wormholes, together with their ergoregions, frame-dragging effects and conditions for horizon avoidance. Second, time-dependent generalisations embedded in cosmological backgrounds of dynamical throats in FLRW or anisotropic spacetimes would allow one to track throat formation and evolution, test energy-condition behaviour during expansion, and assess possible couplings between wormhole dynamics and large-scale cosmic evolution. Third, introducing additional fields or nonminimal couplings (for example, scalar fields coupled to torsion or to the dRGT sector) may produce scalarised or hairy configurations with novel regularity and asymptotic properties; exploring this space requires careful attention to stability and to ghost-free parameter choices. Complementary studies should include linear and nonlinear stability analyses (quasinormal modes and time evolutions), detailed parameter representation and numerical relativity constructions, and investigations of observational signatures (lensing, shadows, and possible gravitational-wave imprints) to identify phenomenologically viable regions of the model space.

\section*{Acknowledgments}
The authors extend their appreciation to the Deanship of Research and  Graduate Studies at King Khalid University for funding this work through a Large Research Project under grant number RGP2/297/46.  The author, S. K. Maurya, is thankful to the administration of the University of Nizwa for their continuous support 
and encouragement for the research work.


\begin{thebibliography}{999}

 
\bibitem{CANTATA:2021asi}
E.~N.~Saridakis \textit{et al.},
 Modified Gravity and Cosmology. An Update by the CANTATA Network,
 \href{https://link.springer.com/book/10.1007/978-3-030-83715-0}{Springer, 
2021}
 \href{https://arxiv.org/abs/2105.1258}{[arXiv:2105.1258 [gr-qc]]}.
 
 


	
	\bibitem{Aldrovandi_Pereira2013}
	Aldrovandi, R. \& Pereira, J.G. {Teleparallel Gravity, An Introduction}, 
\href{https://link.springer.com/book/10.1007/978-94-007-5143-9}{Springer, 2013}.


    \bibitem{Cai:2015emx}
Y.~F.~Cai, S.~Capozziello, M.~De Laurentis and E.~N.~Saridakis,
 f(T) teleparallel gravity and cosmology, 
  \href{https://iopscience.iop.org/article/10.1088/0034-4885/79/10/106901}{{Report Progress Physics} {2016}, {79}, 106901},
 \href{https://arxiv.org/abs/1511.07586}{[arXiv:1511.07586 [gr-qc]]}.
 


	\bibitem{Bahamonde:2021gfp} Bahamonde, S., Dialektopoulos, K., Escamilla-Rivera, C., Farrugia, G., Gakis, V., Hendry, M., Hohmann, M., Said, J.L., Mifsud, J. \& Di Valentino, E., Teleparallel Gravity: From Theory to Cosmology,   \href{https://iopscience.iop.org/article/10.1088/1361-6633/ac9cef}{{Report Progress Physics} {2023}, {86}, 026901}, \href{https://arxiv.org/abs/2106.13793}{[arXiv:2106.13793 [gr-qc]]}.

    \bibitem{Lucas_Obukhov_Pereira2009} Lucas, T.G., Obukhov, Y. \& Pereira, J.G., Regularizing role of teleparallelism, \href{https://journals.aps.org/prd/abstract/10.1103/PhysRevD.80.064043}{{Physical Review D} {2009}, {80}, 064043}, \href{https://arxiv.org/abs/0909.2418}{[arXiv:0909.2418 [gr-qc]]}.
    
	\bibitem{Krssak:2018ywd} Krssak, M., van den Hoogen, R., Pereira, J., Boehmer, C. \& Coley, A., Teleparallel Theories of Gravity: Illuminating a Fully Invariant Approach, \href{https://iopscience.iop.org/article/10.1088/1361-6382/ab2e1f}{{Classical and Quantum Gravity} {2019}, {36}, 183001}, \href{https://arxiv.org/abs/1810.12932}{[arXiv:1810.12932 [gr-qc]]}.


  

     \bibitem{Coley:2019zld} Coley, {A.A.}; van den Hoogen, {R.J.}; McNutt, D.D. Symmetry and Equivalence in Teleparallel Gravity. \href{https://pubs.aip.org/aip/jmp/article-abstract/61/7/072503/395706/Symmetry-and-equivalence-in-teleparallel-gravity?redirectedFrom=fulltext}{{Journal of Mathematical Physics} {2020}, {61}, 072503}, 
\href{https://arxiv.org/abs/1911.03893}{[arXiv:1911.03893 [gr-qc]]}.
	
	\bibitem{Krssak_Pereira2015} Krssak, M. \& Pereira, J.G., Spin Connection and Renormalization of Teleparallel Action, \href{https://link.springer.com/article/10.1140/epjc/s10052-015-3749-2}{{The European Physical Journal C} {2015}, {75}, 519}, \href{https://arxiv.org/abs/1504.07683}{[arXiv:1504.07683 [gr-qc]]}.

	\bibitem{hohmannfq} Hohmann, M. General covariant symmetric teleparallel cosmology. \href{https://journals.aps.org/prd/abstract/10.1103/PhysRevD.104.124077}{\textit{Physical Review D} \textbf{2021}, \textit{104}, 124077}. 
	\href{https://arxiv.org/abs/2109.01525}{[arXiv:2109.01525 [gr-qc]]}

\bibitem{chinea1988symmetries} Chinea, F., Symmetries in tetrad theories, \href{https://iopscience.iop.org/article/10.1088/0264-9381/5/1/018}{\textit{Classical and Quantum Gravity} \textbf{1988}, \textit{5}, 135}.
	
	\bibitem{estabrook1996moving} Estabrook, F. \& Wahlquist, H., Moving frame formulations of 4-geometries having isometries, \href{https://iopscience.iop.org/article/10.1088/0264-9381/13/6/008}{\textit{Classical and Quantum Gravity} \textbf{1996}, \textit{13}, 1333 }.
	
	\bibitem{papadopoulos2012locally} Papadopoulos, G. \& Grammenos, T., Locally homogeneous spaces, induced Killing vector fields and applications to Bianchi prototypes, \href{https://pubs.aip.org/aip/jmp/article/53/7/072502/94967/Locally-homogeneous-spaces-induced-Killing-vector}{\textit{Journal of Mathematical Physics} \textbf{2012}, \textit{53}, 072502 }, \href{https://arxiv.org/abs/1106.3897}{[arXiv:1106.3897 [gr-qc]]} .

 	\bibitem{olver1995equivalence} Olver, P. \textit{Equivalence, invariants and symmetry},  \href{https://books.google.ca/books/about/Equivalence_Invariants_and_Symmetry.html?id=YuTzf61HILAC&redir_esc=y}{Cambridge University Press, 1995}.
	
	\bibitem{MCH} McNutt, D.D., Coley, A.A. \& van den Hoogen, R.J., A frame based approach to computing symmetries with non-trivial isotropy groups, \href{https://pubs.aip.org/aip/jmp/article/64/3/032503/2881713/A-frame-based-approach-to-computing-symmetries}{{Journal of Mathematical Physics} {2023}, {64}, 032503}, \href{https://arxiv.org/abs/2302.11493}{[arXiv:2302.11493 [gr-qc]]}.

	
	
	
	
\bibitem{Chen:2010va}
S.~H.~Chen, J.~B.~Dent, S.~Dutta and E.~N.~Saridakis,
 Cosmological perturbations in f(T) gravity,   
\href{https://journals.aps.org/prd/abstract/10.1103/PhysRevD.83.023508}{\textit{
Physical Review D} 2011, 83, 023508}. \href{https://arxiv.org/abs/1008.1250} 
{[arXiv:1008.1250 [astro-ph.CO]]}.

  


 	\bibitem{Ferraro:2008ey} Ferraro, R. \& Fiorini, F. On Born-Infeld Gravity 
in Weitzenbock spacetime, 
\href{https://journals.aps.org/prd/abstract/10.1103/PhysRevD.78.124019}{\textit{
Physical Review D} \textbf{2008}, \textit{78}, 124019}, 
\href{https://arxiv.org/abs/0812.1981}{[arXiv:0812.1981 [gr-qc]]}.   

  
  %\cite{Krssak:2015oua}
\bibitem{Krssak:2015oua}
M.~Kr{\v{s}}{\v{s}}{\'a}k and E.~N.~Saridakis,
 The covariant formulation of f(T) gravity, 
\href{https://iopscience.iop.org/article/10.1088/0264-9381/33/11/115009}{Class. 
Quantum Grav. \textbf{33}, no.11, 115009 (2016} 
\href{https://arxiv.org/abs/1510.08432}{[arXiv:1510.08432 [gr-qc]]}. 
  
  
  
	
    \bibitem{golov3} Golovnev, A. \& Guzman, M.-J., Bianchi identities in 
$f(T)$-gravity: Paving the way to confrontation with astrophysics, 
\href{
https://www.sciencedirect.com/science/article/pii/S0370269320306092?via%3Dihub}{
{Physics Letter B} {2020}, {810}, 135806}, 
\href{https://arxiv.org/abs/2006.08507}{[ArXiv:2006.08507 [gr-qc]]}.
  
  
   
  
  
  
  
	\bibitem{kayashi} Hayashi, K.;  Shirafuji, T. New general relativity, \href{https://journals.aps.org/prd/abstract/10.1103/PhysRevD.19.3524}{\textit{Physical Review D} \textbf{1979}, \textit{19}, 3524}.
	
	\bibitem{beltranngr} Jimenez, J.B.;  Dialektopoulos, K.F. Non-Linear Obstructions for Consistent New General Relativity. \href{https://iopscience.iop.org/article/10.1088/1475-7516/2020/01/018}{\textit{Journal of Cosmology and Astroparticle Physics} \textbf{2020}, \emph{2020}, 018}. 
	\href{https://arxiv.org/abs/1907.10038}{[arXiv:1907.10038 [gr-qc]]}.
	
	\bibitem{bahamondengr} Bahamonde, S.; Blixt, D.; Dialektopoulos, K.F.;  Hell A. Revisiting Stability in New General Relativity. \href{https://journals.aps.org/prd/abstract/10.1103/PhysRevD.111.064080}{\textit{Physical Review D} 2025, 111, 064080}. \href{https://arxiv.org/abs/2404.02972} {[arXiv:2404.02972 [gr-qc]]}.
	
	\bibitem{heisenberg1} Heisenberg, L. Review on $f(Q)$ Gravity. \href{https://www.sciencedirect.com/science/article/abs/pii/S0370157324000516}{\textit{Physics Report} 2024, 1066, 1-78}. \href{https://arxiv.org/abs/2309.15958}{[arXiv:2309.15958 [gr-qc]]}.
	
	\bibitem{heisenberg2} Heisenberg, L.; Hohmann, M.;  Kuhn, S. Cosmological teleparallel perturbations. \href{https://iopscience.iop.org/article/10.1088/1475-7516/2024/03/063}{Journal of Cosmology and Astroparticle Physics, 2024, 03, 063}. \href{https://arxiv.org/abs/2311.05495}{[arXiv:2311.05495 [gr-qc]]}.
	
	\bibitem{faithman1} Flathmann, K.;  Hohmann, M. Parametrized post-Newtonian limit of generalized scalar-nonmetricity theories of gravity. \href{https://journals.aps.org/prd/abstract/10.1103/PhysRevD.105.044002}{\textit{Physical Review D} \textbf{2022}, \textit{105}, 044002}, \href{https://arxiv.org/abs/2111.02806}{[arXiv:2111.02806 [gr-qc]]}.
	
	
	\bibitem{jimeneztrinity} Jimenez, J.B.; Heisenberg, L.;  Koivisto, T.S. The Geometrical Trinity of Gravity. \href{https://www.mdpi.com/2218-1997/5/7/173}{\textit{Universe} \textbf{2019}, \textit{5}, 173}. \href{https://arxiv.org/abs/1903.06830}{[arXiv:1903.06830 [gr-qc]]}.
	
	\bibitem{nakayama} Nakayama, Y. Geometrical trinity of unimodular gravity. \href{https://iopscience.iop.org/article/10.1088/1361-6382/acd100}{\textit{Classical and Quantum Gravity} \textbf{2023}, \textit{40}, 125005}. 
	\href{https://arxiv.org/abs/2209.09462}{[arXiv:2209.09462 [gr-qc]]}.
	
	\bibitem{ftqgravity} Xu, Y.; Li, G.; Harko, T.;  Liang, S.-D. $f(Q,T)$ gravity. \href{https://link.springer.com/article/10.1140/epjc/s10052-019-7207-4}{\textit{The European Physical Journal C} \textbf{2019}, \textit{79}, 708}. \href{https://arxiv.org/abs/1908.04760}{[arXiv:1908.04760 [gr-qc]]}.
	
	\bibitem{frqspecial} Maurya, D.C.; Yesmakhanova, K.; Myrzakulov, R.;  Nugmanova, G. Myrzakulov, FLRW Cosmology in Metric-Affine $F(R,Q)$ Gravity. \href{https://iopscience.iop.org/article/10.1088/1674-1137/ad6e62}{\textit{Chinese Physics C} \textbf{2024}, 48, 12}. \href{https://arxiv.org/abs/2403.11604}{[arXiv:2403.11604 [gr-qc]]}.
	
	\bibitem{frtspecial} Maurya, D.C.;  Myrzakulov, R. Exact Cosmology in Myrzakulov Gravity. \href{https://link.springer.com/article/10.1140/epjc/s10052-024-12983-4}{\textit{The European Physical Journal C} \textbf{2024}, \textit{84}, 625}, \href{https://arxiv.org/abs/2402.02123}{[arXiv:2402.02123 [gr-qc]]}.
	
	\bibitem{frttheory} Harko, T.; Lobo, F.S.N.; Nojiri, S.;  Odintsov, S.D. $f(R,T)$ gravity. \href{https://journals.aps.org/prd/abstract/10.1103/PhysRevD.84.024020}{\textit{Physical Review D} \textbf{2011}, \textit{84}, 024020.}% 
	\href{https://arxiv.org/abs/1104.2669}{[arXiv:1104.2669 [gr-qc]]}.

    \bibitem{Saridakis:2019qwt}
     E.~N.~Saridakis, S.~Myrzakul, K.~Myrzakulov and K.~Yerzhanov,
 Cosmological applications of $F(R,T)$ gravity with dynamical curvature and torsion.
  \href{https://journals.aps.org/prd/abstract/10.1103/PhysRevD.102.023525}{\textit{Physical Review D} \textbf{2020}, \textit{102}, 023525}. \href{https://arxiv.org/abs/1912.03882}{[arXiv:1912.03882 [gr-qc]]}.

   
%\cite{Anagnostopoulos:2021ydo}
\bibitem{Anagnostopoulos:2021ydo}
F.~K.~Anagnostopoulos, S.~Basilakos and E.~N.~Saridakis,
First evidence that non-metricity f(Q) gravity could challenge 
$\Lambda$CDM,
\href{ 
https://www.sciencedirect.com/science/article/10.1016/j.physletb.2021.136634}{
\textit Phys. Lett. B \textbf{822}, 136634 (2021)} 
\href{https://arxiv.org/abs/2104.15123}{[arXiv:2104.15123 [gr-qc]]}.


 

 %\cite{Basilakos:2025olm}
\bibitem{Basilakos:2025olm}
S.~Basilakos, A.~Paliathanasis and E.~N.~Saridakis,
 Equivalence of f(Q) cosmology with quintom-like scenario: The phantom field 
as effective realization of the non-trivial connection, 
\href{
https://www.sciencedirect.com/science/article/10.1016/j.physletb.2025.139658}{
\textit Phys. Lett. B \textbf{868}, 139658 (2025)} 
\href{https://arxiv.org/abs/2503.19864}{[arXiv:2503.19864 [gr-qc]]}.






  \bibitem{derham1} de Rham, C., Gabadadze, G., \& Tolley, A.J. Resummation of 
Massive Gravity. 
\href{https://journals.aps.org/prl/abstract/10.1103/PhysRevLett.106.231101}{{
Phys. Rev. Lett.} 106, 231101 (2011)}, 
\href{https://arxiv.org/abs/1011.1232}{[arXiv:1011.1232 [gr-qc]]}.

    \bibitem{derham2} de Rham, C., Massive Gravity, 
\href{https://link.springer.com/article/10.12942/lrr-2014-7}{Living Rev. 
Relativity (2014), 17, 7}, 
\href{https://arxiv.org/abs/1401.4173}{[arXiv:1401.4173 [gr-qc]]}.

    \bibitem{derham3} de Rham, C., Hinterbichler, K., Johnson, L.A., On the 
(A)dS Decoupling Limits of Massive Gravity, 
\href{https://link.springer.com/article/10.1007/JHEP09(2018)154}{JHEP (2018), 
09, 154}, \href{https://arxiv.org/abs/1807.08754}{[arXiv:1807.08754 [hep-th]]}.

    \bibitem{derham4} de Rham, C., Melville, S. \& Tolley, A.J. , Improved 
Positivity Bounds and Massive Gravity, 
\href{https://link.springer.com/article/10.1007/JHEP04(2018)083}{JHEP (2018), 
04, 083}, \href{https://arxiv.org/abs/1710.09611}{[arXiv:1710.09611 [hep-th]]}.

    \bibitem{derham5} de Rham, C., Melville, S., Tolley, A.J. \& Zhou, S.-Y., 
Massive Galileon Positivity Bounds, 
\href{https://link.springer.com/article/10.1007/JHEP09(2017)072}{JHEP (2017), 
09, 072}, \href{https://arxiv.org/abs/1702.08577}{[arXiv:1702.08577 [hep-th]]}.

    \bibitem{derham6} de Rham, C., Heisenberg, L. \& Ribeiro, R.H., On couplings 
to matter in massive (bi-)gravity, 
\href{https://iopscience.iop.org/article/10.1088/0264-9381/32/3/035022}{Class. 
Quantum Grav. (2015), 32, 035022}, 
\href{https://arxiv.org/abs/1408.1678}{[arXiv:1408.1678 [hep-th]]}.

    \bibitem{massivefrgravity1} Bhar, P., Properties of the wormhole model in 
de 
Rham–Gabadadze–Tolley like massive gravity with specific matter density, 
\href{https://doi.org/10.1140/epjc/s10052-024-13670-0}{{The European Physical 
Journal C} {(2025)}, {85}}, 
\href{https://arxiv.org/abs/2408.02717}{[ArXiv:2408.02717 [gr-qc]]}.











 


	
    \bibitem{golov1} Golovnev, A.; Guzman, M.-J. Approaches to spherically symmetric solutions in $f(T)$-gravity. \href{https://www.mdpi.com/2218-1997/7/5/121}{\textit{Universe} \textbf{2021}, \textit{7}, 121}. \href{https://arxiv.org/abs/2103.16970}{[arXiv:2103.16970 [gr-qc]]}.
	
    \bibitem{golov2} Golovnev, A. Issues of Lorentz-invariance in $f(T)$-gravity and calculations for spherically symmetric solutions. \href{https://iopscience.iop.org/article/10.1088/1361-6382/ac2136}{\textit{Classical and Quantum Gravity} \textbf{2021}, \textit{38}, 197001}. \href{https://arxiv.org/abs/2105.08586}{[arXiv:2105.08586 [gr-qc]]}.
	
	\bibitem{debenedictis} DeBenedictis, A.; Iliji\'c, S.; Sossich, M. On spherically symmetric vacuum solutions and horizons in covariant $f(T)$ gravity theory. \href{https://journals.aps.org/prd/abstract/10.1103/PhysRevD.105.084020}{\textit{Physical Review D} \textbf{2022}, \textit{105}, 084020}. \href{https://arxiv.org/abs/2202.08958}{[arXiv:2202.08958 [gr-qc]]}.
	
	\bibitem{baha1} Bahamonde, S.; Camci, U. Exact Spherically Symmetric Solutions in Modified Teleparallel gravity. \href{https://www.mdpi.com/2073-8994/11/12/1462}{\textit{Symmetry} \textbf{2019}, \textit{11}, 1462}. \href{https://arxiv.org/abs/1911.03965v2}{[arXiv:1911.03965 [gr-qc]]}.
	
	\bibitem{awad1} Awad, A.; Golovnev, A.; Guzman, M.-J.; El Hanafy, W. Revisiting diagonal tetrads: New Black Hole solutions in $f(T)$-gravity. \href{https://link.springer.com/article/10.1140/epjc/s10052-022-10939-0}{\textit{The European Physical Journal C} \textbf{2022}, \textit{82}, 972}. \href{https://arxiv.org/abs/2207.00059}{[arXiv:2207.00059 [gr-qc]]}.
	
	\bibitem{bahagolov1} Bahamonde, S.; Golovnev, A.; Guzm\'an, M.-J.; Said, J.L.; Pfeifer, C. Black Holes in $f(T,B)$ Gravity: Exact and Perturbed Solutions. \href{https://iopscience.iop.org/article/10.1088/1475-7516/2022/01/037}{\textit{J. Cosmol. Astropart. Phys.} \textbf{2022}, \textit{1}, 037}. \href{https://arxiv.org/abs/2110.04087}{[arXiv:2110.04087 [gr-qc]]}.
	
	\bibitem{baha6} Bahamonde, S.; Faraji, S.; Hackmann, E.; Pfeifer, C. Thick accretion disk configurations in the Born-Infeld teleparallel gravity. \href{https://journals.aps.org/prd/abstract/10.1103/PhysRevD.106.084046}{\textit{Physical Review D} \textbf{2022}, \textit{106}, 084046}. \href{https://arxiv.org/abs/2209.00020}{[arXiv:2209.00020 [gr-qc]]}.
	
	\bibitem{nashed5} Nashed, G.G.L. Quadratic and cubic spherically symmetric black holes in the modified teleparallel equivalent of general relativity: Energy and thermodynamics. \href{https://iopscience.iop.org/article/10.1088/1361-6382/abf89b}{\textit{Classical and Quantum Gravity} \textbf{2021}, \textit{38}, 125004}. \href{https://arxiv.org/abs/2105.05688}{[arXiv:2105.05688 [gr-qc]]}.
	
	\bibitem{pfeifer2} Pfeifer, C.; Schuster, S. Static spherically symmetric black holes in weak $f(T)$-gravity. \href{https://www.mdpi.com/2218-1997/7/5/153}{\textit{Universe} \textbf{2021}, \textit{7}, 153}. \href{https://arxiv.org/abs/2104.00116v2}{[arXiv:2104.00116 [gr-qc]]}.
	
	\bibitem{elhanafy1} El Hanafy, W.; Nashed, G.G.L. Exact Teleparallel Gravity of Binary Black Holes, \href{https://link.springer.com/article/10.1007/s10509-016-2662-y}{\textit{Astrophysical and Space Science} \textbf{2016}, \textit{361}, 68}. \href{https://arxiv.org/abs/1507.07377}{[arXiv:1507.07377 [gr-qc]]}.
	
	\bibitem{benedictis3} Aftergood, J.; DeBenedictis, A. Matter Conditions for Regular Black Holes in $f(T)$ Gravity. \href{https://journals.aps.org/prd/abstract/10.1103/PhysRevD.90.124006}{\textit{Physical Review D} \textbf{2014}, \textit{90}, 124006}, \href{https://arxiv.org/abs/1409.4084}{[arXiv:1409.4084 [gr-qc]]}.
	
	\bibitem{baha10} Bahamonde, S.; Doneva, D.D.; Ducobu, L.; Pfeifer, C.; Yazadjiev, S.S. Spontaneous Scalarization of Black Holes in Gauss-Bonnet Teleparallel Gravity, \href{https://journals.aps.org/prd/abstract/10.1103/PhysRevD.107.104013}{\textit{Physical Review D} \textbf{2023}, \textit{107}, 104013}. \href{https://arxiv.org/abs/2212.07653}{[arXiv:2212.07653 [gr-qc]]}.
	
	\bibitem{baha4} Bahamonde, S.; Ducobu, L.; Pfeifer, C. Scalarized Black Holes in Teleparallel Gravity. \href{https://iopscience.iop.org/article/10.1088/1475-7516/2022/04/018}{\textit{J. Cosmol. Astropart. Phys.} \textbf{2022}, {\textit{2022}}, 018}. \href{https://arxiv.org/abs/2201.11445v2}{[arXiv:2201.11445 [gr-qc]]}.

	\bibitem{ruggiero1} Ruggiero, M.L. \& Radicella, N., Weak-Field Spherically Symmetric Solutions in $f(T)$ gravity, \href{https://journals.aps.org/prd/abstract/10.1103/PhysRevD.91.104014}{\textit{Physical Review D} \textbf{2015}, \textit{91}, 104014}, \href{https://arxiv.org/abs/1501.02198}{[arXiv:1501.02198 [gr-qc]]}.
	  
	\bibitem{ruggiero2} Iorio, L.; Radicella, N.;  Ruggiero, M.L. Constraining $f(T)$ gravity in the Solar System. \href{https://iopscience.iop.org/article/10.1088/1475-7516/2015/08/021}{\textit{J. Cosmol. Astropart. Phys.} \textbf{2015}, \textit{{2015}}, 021}, \href{https://arxiv.org/abs/1505.06996}{[arXiv:1505.06996 [gr-qc]]}.
	
	\bibitem{sahoo1} Pradhan, S.; Bhar, P.; Mandal, S.; Sahoo, P.K.; Bamba, K. The Stability of Anisotropic Compact Stars Influenced by Dark Matter under Teleparallel Gravity: An Extended Gravitational Deformation Approach. \href{https://link.springer.com/article/10.1140/epjc/s10052-025-13849-z}{\textit{The European Physical Journal C} \textbf{2025}, \textit{85}, 127}. \href{https://arxiv.org/abs/2408.03967}{[arXiv:2408.03967 [gr-qc]]}.
	
	\bibitem{sahoo2} Mohanty, D.; Ghosh, S.; Sahoo, P.K. Charged gravastar model in noncommutative geometry under $f(T)$ gravity. \href{https://www.sciencedirect.com/science/article/abs/pii/S2212686424002747}{\textit{Phys. Dark Universe} \textbf{2025}, \textit{46}, 101692}. \href{https://arxiv.org/abs/2410.05679}{[arXiv:2410.05679 [gr-qc]]}.
	
	\bibitem{calza} Calza, M.;  Sebastiani, L. A class of static spherically symmetric solutions in $f(T)$-gravity. \href{https://link.springer.com/article/10.1140/epjc/s10052-024-12801-x}{\textit{{The European Physical Journal C}} \textbf{2024}, \textit{84}, 476.} 
	\href{https://arxiv.org/abs/2309.04536}{[arXiv:2309.04536 [gr-qc]]}.

	\bibitem{pfeifer2022quick} Pfeifer, C., A quick guide to spacetime symmetry and symmetric solutions in teleparallel gravity, \textbf{2022}, \href{https://arxiv.org/abs/2201.04691}{[arXiv:2201.04691 [gr-qc]]}.

	\bibitem{SSpaper} Coley, A.A., Landry, A., van den Hoogen, R.J. \& McNutt, D.D., Spherically symmetric teleparallel geometries, \href{https://link.springer.com/article/10.1140/epjc/s10052-024-12629-5}{{The European Physical Journal C} {2024}, {84}, 334}, \href{https://arxiv.org/abs/2402.07238}{[ArXiv:2402.07238 [gr-qc]]}.

    
	\bibitem{nonvacSSpaper} Landry, A., Static spherically symmetric perfect fluid solutions in teleparallel F(T) gravity, \href{https://www.mdpi.com/2075-1680/13/5/333}{{Axioms} {2024}, {13} (5), 333}, \href{https://arxiv.org/abs/2405.09257}{[arXiv:2405.09257 [gr-qc]]}.

    \bibitem{scalarfieldSSpaper} Landry, A., Scalar Field Static Spherically Symmetric Solutions in Teleparallel $F(T)$ Gravity, \href{https://www.mdpi.com/2227-7390/13/6/1003}{{Mathematics} {2025}, {13} (6), 1003}, \href{https://arxiv.org/abs/2503.14465}{[arXiv:2503.14465 [gr-qc]]}
		
	\bibitem{roberthudsonSSpaper} van den Hoogen, R.J. \& Forance, H., Teleparallel Geometry with Spherical Symmetry: The diagonal and proper frames, \href{https://iopscience.iop.org/article/10.1088/1475-7516/2024/11/033}{{Journal of Cosmology and Astrophysics} {2024}, {11}, 033}, \href{https://arxiv.org/abs/2408.13342}{[arXiv:2408.13342 [gr-qc]]}

    \bibitem{scalarfieldKSpaper} Landry, A., Scalar field Kantowski-Sachs spacetime solutions in teleparallel $F(T)$ gravity, \href{https://www.mdpi.com/2218-1997/11/1/26}{\textit{Universe} \textbf{2025}, 11(1), 26}, \href{https://arxiv.org/abs/2501.11160}{[arXiv:2501.11160 [gr-qc]]}.
	

    \bibitem{nonvacKSpaper} Landry, A., Kantowski-Sachs spherically symmetric solutions in teleparallel $F(T)$ gravity, \href{https://www.mdpi.com/2073-8994/16/8/953}{\textit{Symmetry} \textbf{2024} \textit{16} (8), 953}, \href{https://arxiv.org/abs/2406.18659}{[arXiv:2406.18659 [gr-qc]]}.
    
	\bibitem{TdSpaper} Coley, A.A., Landry, A., van den Hoogen, R.J. \& McNutt, D.D., Generalized Teleparallel de Sitter geometries, \href{https://link.springer.com/article/10.1140/epjc/s10052-023-12150-1}{{The European Physical Journal C} {2023}, {83}, 977}, \href{https://arxiv.org/abs/2307.12930}{[ArXiv:2307.12930 [gr-qc]]}. 

	
	\bibitem{sharif2009teleparallel} Sharif, M. \& Majeed, B., Teleparallel Killing Vectors of Spherically Symmetric Spacetimes, \href{https://iopscience.iop.org/article/10.1088/0253-6102/52/3/11}{\textit{Communications In Theoretical Physics} \textbf{2009}, \textit{52}, 435}, \href{https://arxiv.org/abs/0905.3212}{[arXiv:0905.3212 [gr-qc]]}.
		
  	\bibitem{attractor} Duchaniya, L.K., Gandhi, K. \&  Mishra, B., Attractor behavior of $f(T)$ modified gravity and the cosmic acceleration, \href{https://www.sciencedirect.com/science/article/pii/S2212686424000438?via%3Dihub}{\textit{Physics of the Dark Universe} \textbf{2024}, \textit{44}, 101464}, \href{https://arxiv.org/abs/2303.09076}{[arXiv:2303.09076 [gr-qc]]}.
 
  	
  	
  	
  	
  	 
 
%\cite{Gonzalez:2011dr}
\bibitem{Gonzalez:2011dr}
P.~A.~Gonzalez, E.~N.~Saridakis and Y.~Vasquez,
 Circularly symmetric solutions in three-dimensional Teleparallel, f(T) and 
Maxwell-f(T) gravity,
\href{https://link.springer.com/article/10.1007/JHEP07(2012)053}{JHEP 
\textbf{07}, 053 (2012)}, 
\href{https://arxiv.org/abs/1110.4024}{[arXiv:1110.4024 [gr-qc]]}.

 
 

%\cite{Ferraro:2011ks}
\bibitem{Ferraro:2011ks}
R.~Ferraro and F.~Fiorini,
 Spherically symmetric static spacetimes in vacuum f(T) gravity, 
 \href{https://journals.aps.org/prd/abstract/10.1103/PhysRevD.84.083518}{Phys. 
Rev. D \textbf{84}, 083518 (2011)}, 
 \href{https://arxiv.org/abs/1109.4209}{[arXiv:1109.4209 [gr-qc]]}.

 

%\cite{Capozziello:2012zj}
\bibitem{Capozziello:2012zj}
S.~Capozziello, P.~A.~Gonzalez, E.~N.~Saridakis and Y.~Vasquez,
 Exact charged black-hole solutions in D-dimensional f(T) gravity: torsion vs 
curvature analysis, 
\href{https://link.springer.com/article/10.1007/JHEP02(2013)039}{JHEP 
\textbf{02}, 039 (2013)}, 
\href{https://arxiv.org/abs/1110.4024}{[arXiv:1210.1098 [hep-th]]}.

 
 
 

%\cite{Jamil:2012ti}
\bibitem{Jamil:2012ti}
M.~Jamil, D.~Momeni and R.~Myrzakulov,
 Wormholes in a viable f(T) gravity,    
\href{https://link.springer.com/article/10.1140/epjc/s10052-012-2267-8}{Eur. 
Phys. J. C \textbf{73}, 2267 (2013)}, 
\href{https://arxiv.org/abs/1212.6017}{[arXiv:1212.6017 [gr-qc]]}.

 
 
  

%\cite{Nashed:2013bfa}
\bibitem{Nashed:2013bfa}
G.~G.~L.~Nashed,
 Spherically symmetric charged-dS solution in $f(T)$ gravity theories, 
  \href{https://journals.aps.org/prd/abstract/10.1103/PhysRevD.88.104034}{Phys. 
Rev. D \textbf{88}, 104034 (2013)}, 
 \href{https://arxiv.org/abs/1311.3131}{[arXiv:1311.3131 [gr-qc]]}.
 
 
  


%\cite{Nashed:2013owg}
\bibitem{Nashed:2013owg}
G.~G.~L.~Nashed,
 A special exact spherically symmetric solution in f(T) gravity theories,    
\href{https://link.springer.com/article/10.1007/s10714-013-1566-1}{Gen. 
Rel. Grav. \textbf{45}, 1887-1899 (2013)}, 
 \href{https://arxiv.org/abs/1502.05219}{[arXiv:1502.05219 [gr-qc]]}.
 
  
 
%\cite{Mai:2017riq}
\bibitem{Mai:2017riq}
Z.~F.~Mai and H.~Lu,
 Black Holes, Dark Wormholes and Solitons in f(T) Gravities, 
  \href{https://journals.aps.org/prd/abstract/10.1103/PhysRevD.95.124024}{Phys. 
Rev. D \textbf{95}, no.12, 124024 (2017)}, 
 \href{https://arxiv.org/abs/1704.05919}{[arXiv:1704.05919 [hep-th]]}.
 
 
  
	
	\bibitem{salvatore1} Nurbaki, A.N., Capozziello, S. \& Deliduman, C., 
Spherical and cylindrical solutions in $f(T)$ gravity by Noether Symmetry 
Approach, 
\href{https://link.springer.com/article/10.1140/epjc/s10052-020-7666-7}{\textit{
The European Physical Journal C} \textbf{2020}, \textit{80}, 108}, 
\href{https://arxiv.org/abs/2001.02304}{[ArXiv:2001.02304 [gr-qc]]} . 


   
	\bibitem{bahamonde2021exploring} Bahamonde, S., Valcarcel, J., J\:arv, L. 
\& 
Pfeifer, C., Exploring axial symmetry in modified teleparallel gravity, 
\href{https://journals.aps.org/prd/abstract/10.1103/PhysRevD.103.044058}{\textit
{Physical Review D} \textbf{2021}, \textit{103}, 044058}, 
\href{https://arxiv.org/abs/2012.09193}{[arXiv:2012.09193 [gr-qc]]}.
  
 %\cite{Golovnev:2021lki}
\bibitem{Golovnev:2021lki}
A.~Golovnev,
 Issues of Lorentz-invariance in f(T) gravity and calculations for 
spherically 
symmetric solutions, 
 \href{https://iopscience.iop.org/article/10.1088/1361-6382/ac2136}{Class. 
Quant. Grav. \textbf{38}, no.19, 197001 (2021)}, 
\href{https://arxiv.org/abs/2105.08586}{[arXiv:2105.08586 [gr-qc]]}.


 

%\cite{Nashed:2021pah}
\bibitem{Nashed:2021pah}
G.~G.~L.~Nashed and E.~N.~Saridakis,
 Stability of motion and thermodynamics in charged black holes in f(T) 
gravity, 
\href{https://iopscience.iop.org/article/10.1088/1475-7516/2022/05/017}{JCAP 
\textbf{05}, no.05, 017 (2022)}, 
\href{https://arxiv.org/abs/2111.06359}{[arXiv:2111.06359 [gr-qc]]}.



 
 

%\cite{Ren:2021uqb}
\bibitem{Ren:2021uqb}
X.~Ren, Y.~Zhao, E.~N.~Saridakis and Y.~F.~Cai,
 Deflection angle and lensing signature of covariant f(T) gravity, 
 \href{https://iopscience.iop.org/article/10.1088/1475-7516/2021/10/062}{JCAP 
\textbf{10}, 062 (2021)}, 
\href{https://arxiv.org/abs/2105.04578}{[arXiv:2105.04578 [astro-ph.CO]]}.


 



%\cite{Papanikolaou:2022hkg}
\bibitem{Papanikolaou:2022hkg}
T.~Papanikolaou, C.~Tzerefos, S.~Basilakos and E.~N.~Saridakis,
 No constraints for f(T) gravity from gravitational waves induced from 
primordial black hole fluctuations, 
 \href{https://link.springer.com/article/10.1140/epjc/s10052-022-11157-4}{Eur. 
Phys. J. C \textbf{83}, no.1, 31 (2023)}, 
\href{https://arxiv.org/abs/2205.06094}{[arXiv:2205.06094 [gr-qc]]}.

 
 

%\cite{Awad:2022fhx}
\bibitem{Awad:2022fhx}
A.~Awad, A.~Golovnev, M.~J.~Guzm{\'a}n and W.~El Hanafy,
 Revisiting diagonal tetrads: new Black Hole solutions in f(T) gravity, 
  \href{https://link.springer.com/article/10.1140/epjc/s10052-022-10939-0}{Eur. 
Phys. J. C \textbf{82}, no.10, 972 (2022)}, 
\href{https://arxiv.org/abs/2207.00059}{[arXiv:2207.00059 [gr-qc]]}.

 
 
 
  
 

%\cite{Zhao:2022gxl}
\bibitem{Zhao:2022gxl}
Y.~Zhao, X.~Ren, A.~Ilyas, E.~N.~Saridakis and Y.~F.~Cai,
 Quasinormal modes of black holes in f(T) gravity, 
  \href{https://iopscience.iop.org/article/10.1088/1475-7516/2022/10/087}{JCAP 
\textbf{10}, 087 (2022)}, 
\href{https://arxiv.org/abs/2204.11169}{[arXiv:2204.11169 [astro-ph.CO]]}.


 
  

 
%\cite{DeBenedictis:2022sja}
\bibitem{DeBenedictis:2022sja}
A.~DeBenedictis, S.~Iliji{\'c} and M.~Sossich,
 Spherically symmetric vacuum solutions and horizons in covariant f(T) 
gravity 
theory, 
 \href{https://journals.aps.org/prd/abstract/10.1103/PhysRevD.105.084020}{Phys. 
Rev. D \textbf{105}, no.8, 084020 (2022)}, 
 \href{https://arxiv.org/abs/2202.08958}{[arXiv:2202.08958 [gr-qc]]}.
 
      \bibitem{landryvandenhoogen1} Landry, A. \& van den Hoogen, R.J., 
Teleparallel Minkowski Spacetime with Perturbative Approach for Teleparallel 
Theories on the Proper Frame, 
\href{https://www.mdpi.com/2218-1997/9/5/232}{\textit{Universe} \textbf{2013}, 
\textit{9}, 232}, \href{https://arxiv.org/abs/2303.16089}{[arXiv:2303.16089 
[gr-qc]]}.
  
  
  
  

    \bibitem{coleylandrygholami} Coley, A.A.; Landry, A.; Gholami, F. 
Teleparallel Robertson-Walker Geometries and Applications. 
\href{https://www.mdpi.com/2218-1997/9/10/454}{\textit{Universe} \textbf{2023}, 
\textit{9}, 454}, \href{https://arxiv.org/abs/2310.14378}{[arXiv:2310.14378 
[gr-qc]]}.
  



%\cite{Rani:2022tmf}
\bibitem{Rani:2022tmf}
S.~Rani, M.~B.~A.~Sulehri, A.~Jawad and U.~Zafar,
 Casimir traversable wormholes for GUP corrected energy densities in f(T) 
gravity, 
  \href{https://www.worldscientific.com/doi/10.1142/S0219887823500548}{    
Int. J. Geom. Meth. Mod. Phys. \textbf{20} (2022) 03}.

 

%\cite{Wang:2023qfm}
\bibitem{Wang:2023qfm}
Q.~Wang, X.~Ren, B.~Wang, Y.~F.~Cai, W.~Luo and E.~N.~Saridakis,
 Galaxy{\textendash}Galaxy Lensing Data: f(T) Gravity Challenges General 
Relativity, 
 \href{https://iopscience.iop.org/article/10.3847/1538-4357/ad47c0}{
Astrophys. J. \textbf{969}, no.2, 119 (2024)}, 
 \href{https://arxiv.org/abs/2312.17053}{[arXiv:2312.17053 [astro-ph.CO]]}.
 
 
  

%\cite{Addazi:2023pfx}
\bibitem{Addazi:2023pfx}
A.~Addazi and S.~Capozziello,
 Black hole shadow and chaos bound violation in f(T) teleparallel gravity, 
 
\href{
https://www.sciencedirect.com/science/article/pii/S0370269323001624?via3Dihub}{
\textit Phys. Lett. B \textbf{839}, 137828 (2023)} 
\href{https://arxiv.org/abs/2303.01956}{[arXiv:2303.01956 [gr-qc]]}.


 

	

    \bibitem{FTBcosmogholamilandry} Gholami, F.; Landry, A. Cosmological 
solutions in teleparallel $F(T,B)$ gravity. 
\href{https://www.mdpi.com/2073-8994/17/1/60}{\textit{Symmetry} \textbf{2025}, 
\textit{17}, 060},  \href{https://arxiv.org/abs/2411.18455}{[arXiv:2411.18455 
[gr-qc]]}.



%\cite{Wu:2024vcr}
\bibitem{Wu:2024vcr}
C.~Wu, X.~Ren, Y.~Yang, Y.~M.~Hu and E.~N.~Saridakis,
 Background-dependent and classical correspondences between f(Q) and f(T) 
gravity, 
  \href{https://link.springer.com/article/10.1140/epjc/s10052-025-14822-6}{Eur. 
Phys. J. C \textbf{85}, no.10, 1099 (2025)}, 
\href{https://arxiv.org/abs/2412.01104}{[arXiv:2412.01104 [gr-qc]]}.

 
 

    \bibitem{scalarfieldTRW} Landry, A. Scalar field sources Teleparallel 
Robertson-Walker $F(T)$-gravity {solutions.}  
\href{https://www.mdpi.com/2227-7390/13/3/374}{\textit{Mathematics} 
\textbf{2025}, \textit{13}, 374}, 
\href{https://arxiv.org/abs/2501.13895}{[arXiv:2501.13895 [gr-qc]]}.
 
  


 
   


 
  
 
  	
  	
  	
  	
  	
  	
  	
  	
  	
  	
  	
  	
\bibitem{bahabohmer} Bahamonde, S.,  Bohmer, C.G., Carloni, S., Copeland, E.J., Fang, W. \& Tamanini, N., \textit{Dynamical systems applied to cosmology: dark energy and modified gravity}, \href{https://www.sciencedirect.com/science/article/pii/S0370157318302242?via%3Dihub}{Physics Reports \textbf{2018}, \textit{775-777}, 1-122}, \href{https://arxiv.org/abs/1712.03107}{[arXiv:1712.03107 [gr-qc]]}.



\bibitem{tamanini2012tetrads} Tamanini, N. \& Böhmer, C. G., Good and bad tetrads in \(f(T)\) gravity, \href{https://journals.aps.org/prd/abstract/10.1103/PhysRevD.86.044009}{\textit{Physical Review D} \textbf{2012}, \textit{86}, 044009}, \href{https://arxiv.org/abs/1204.4593}{[arXiv:1204.4593 [gr-qc]]}.


\bibitem{bamba2011singularity} K.~Bamba, S.~Nojiri and S.~D.~Odintsov, Time-dependent matter instability and star singularity in $F(R)$ gravity, \href{https://www.sciencedirect.com/science/article/abs/pii/S037026931100311X?via%3Dihub}{\textit{Physics Letters B} \textbf{2011}, \textit{698}, 451-456}, \href{https://arxiv.org/abs/1101.2820}{[arXiv:1101.2820 [gr-qc]]}.



\bibitem{sharif2013wormholes}
Sharif, M. \& Rani, S.,
Wormhole solutions in f(T) gravity with noncommutative geometry, 
\href{https://doi.org/10.1103/PhysRevD.88.123501}{\textit{Physical Review D} 
\textbf{88}, 123501 (2013)}.

\bibitem{davis2015anisotropic} Di Grezia, E., Battista, E., Manfredonia, M. \& Miele, G.,
Spin, torsion, and violation of null energy condition in traversable wormholes, \href{https://doi.org/10.1140/epjp/i2017-11799-6}{
\textit{The European Physical Journal Plus} \textbf{132}, 537 (2017)},
\href{https://arxiv.org/abs/1707.01508}{[arXiv:1707.01508 [gr-qc]]}.

{ 

 \bibitem{Shweta:2023hvm}
Shweta, U.~K.~Sharma and A.~K.~Mishra, Yukawa-Casimir wormholes in 4-D Einstein Gauss-Bonnet gravity, \href{https://www.worldscientific.com/doi/10.1142/S0219887823501402}{
 Int. J. Geom. Meth. Mod. Phys. \textbf{20}(08), 2350140 (2023)}, \href{https://arxiv.org/abs/2004.10750}{[arXiv:2004.10750 [gr-qc]]}.


\bibitem{Gudekli:2022vjs}
E.~G{\"u}dekli, H.~Nazar, G.~Abbas and M.~Saddique,
 Wormholes models in $f (R)$ gravity inspired by non-compact matter source,
\href{https://www.worldscientific.com/doi/10.1142/S0219887822502279}{Int. J. Geom. Meth. Mod. Phys. \textbf{19}(14), 2250227 (2022)}.


\bibitem{Hussain:2022lxb}
I.~Hussain and G.~Mustafa,
Traversable wormholes in Einsteinian-cubic-gravity with hybrid shape functions, \href{https://www.worldscientific.com/doi/abs/10.1142/S0219887822500748}{Int. J. Geom. Meth. Mod. Phys. \textbf{19}(05), 2250074 (2022)} \href{https://arxiv.org/pdf/1903.10907}{[arXiv:1903.10907 [gr-qc]]}.

\bibitem{Abbas:2023jhi}
G.~Abbas, S.~Taj, A.~Siddiqa and Z.~Arbab,
Energy constraints for static wormholes in $f (R,T)$ gravity, \href{https://www.worldscientific.com/doi/10.1142/S0219887823502365}{
Int. J. Geom. Meth. Mod. Phys. \textbf{20}(13), 2350236 (2023)}, \href{https://arxiv.org/abs/2208.08854}{[arXiv:2208.08854 [gr-qc]]}.


\bibitem{Battista:2024gud}
E.~Battista, S.~Capozziello and A.~Errehymy,
Generalized uncertainty principle corrections in Rastall{\textendash}Rainbow Casimir wormholes, \href{https://link.springer.com/article/10.1140/epjc/s10052-024-13656-y}{
Eur. Phys. J. C \textbf{84}(12), 1314 (2024)}, \href{https://arxiv.org/abs/2409.09750}{[arXiv:2409.09750 [gr-qc]]}.
%doi:10.1140/epjc/s10052-024-13656-y

\bibitem{DeFalco:2021ksd}
V.~De Falco, E.~Battista, S.~Capozziello and M.~De Laurentis,
Reconstructing wormhole solutions in curvature based Extended Theories of Gravity, \href{https://link.springer.com/article/10.1140/epjc/s10052-021-08958-4}{Eur. Phys. J. C \textbf{81}(2), 157 (2021)}, \href{https://arxiv.org/abs/2102.01123}{[arXiv:2102.01123 [gr-qc]]}.
%doi:10.1140/epjc/s10052-021-08958-4

\bibitem{DeFalco:2021klh}
V.~De Falco, E.~Battista, S.~Capozziello and M.~De Laurentis,
Testing wormhole solutions in extended gravity through the Poynting-Robertson effect, \href{https://journals.aps.org/prd/abstract/10.1103/PhysRevD.103.044007}{
Phys. Rev. D \textbf{103}(4), 044007 (2021)}, \href{https://arxiv.org/abs/2101.04960}{[arXiv:101.04960 [gr-qc]]}.
%doi:10.1103/PhysRevD.103.044007

\bibitem{Zubair:2023dbe}
M.~Zubair, M.~Farooq, E.~Gudekli, H.~R.~Kousar and G.~D.~A.~Yildiz,
New traversable wormhole solutions in Einstein Gauss{\textendash}Bonnet gravity, 
\href{https://www.worldscientific.com/doi/10.1142/S0219887823501918}{Int. J. Geom. Meth. Mod. Phys. \textbf{20}(11), 2350191 (2023)}.
%doi:10.1142/S0219887823501918

\bibitem{Ahmed:2022myl}
R.~Ahmed, G.~Abbas, H.~Nazar and K.~Iqbal,
Existence of traversable wormholes with varied shape functions in $f(\mathcal{R}^2,{\mathcal{T}} )$ gravity, \href{https://www.worldscientific.com/doi/10.1142/S0219887822501705}{
Int. J. Geom. Meth. Mod. Phys. \textbf{19}(12), 2250170 (2022)}.
%doi:10.1142/S0219887822501705



\bibitem{Ashraf:2023bfg}
A.~Ashraf, S.~Mumtaz, F.~Javed and Z.~Zhang,
Viable embedded wormholes and energy conditions in $f({\mathcal{R}},\mathcal{G})$ gravity, \href{https://www.worldscientific.com/doi/10.1142/S0219887824502086}{
Int. J. Geom. Meth. Mod. Phys. \textbf{21}(12), 2450208 (2024)}, \href{https://arxiv.org/abs/2304.06256}{[arXiv:2304.06256 [gr-qc]]}.
%doi:10.1142/S0219887824502086

\bibitem{Naz:2022smd}
T.~Naz, G.~Mustafa and M.~Farasat Shamir,
Existence of wormholes in $f(\mathcal G)$ gravity using symmetries, \href{https://www.worldscientific.com/doi/10.1142/S0219887822501006}{
Int. J. Geom. Meth. Mod. Phys. \textbf{19}(07), 2250100 (2022)}, \href{https://arxiv.org/abs/2304.06365}{[arXiv:2304.06365 [gr-qc]]}.
%doi:10.1142/S0219887822501006

}
   

    \bibitem{morristhorne1} Morris, M.S. \& Thorne, K.S., Wormholes in spacetime and their use for interstellar travel: A tool for teaching general relativity, \href{https://pubs.aip.org/aapt/ajp/article-abstract/56/5/395/1044276/Wormholes-in-spacetime-and-their-use-for?redirectedFrom=fulltext}{American Journal of Physics (1988), 56, 395}.

    \bibitem{morristhorne2} Morris, M.S., Thorne, K.S. \&  Yurtsever, U., Wormholes, Time Machines, and the Weak Energy Condition, 
    \href{https://journals.aps.org/prl/abstract/10.1103/PhysRevLett.61.1446}{Physical Review Letters, (1988), 61, 1446}.

  
 
    \bibitem{Kontou:2020bta} Kontou, E.-A. \& Sanders, K., Energy conditions in general relativity and quantum field theory, \href{https://iopscience.iop.org/article/10.1088/1361-6382/ab8fcf}{Classical and Quantum Gravity 37, 193001 (2020)}, \href{https://arxiv.org/abs/2003.01815}{[ArXiv:2003.01815 [gr-qc]]}.




  
    
\end{thebibliography}
\end{document}